\documentclass[a4paper,11pt]{article}
\usepackage{caption}
\usepackage{placeins}
\usepackage{float}
\usepackage{jcappub}
\usepackage{lineno}
\usepackage{fontawesome}
\usepackage{orcidlink}
\usepackage{subcaption}
\usepackage{enumitem}
\usepackage[normalem]{ulem}
\usepackage{xcolor}
\usepackage{CJKutf8}

\definecolor{r}{HTML}{EE6677}
\definecolor{b}{HTML}{4477AA}
\definecolor{g}{HTML}{228833}
\definecolor{y}{HTML}{CCBB44}
\definecolor{k}{HTML}{BBBBBB}
\definecolor{p}{HTML}{AA3377}
\definecolor{c}{HTML}{66CCEE}

\usepackage{xspace}

\newcommand{\Om}{\ensuremath{\Omega_m}\xspace}
\newcommand{\Ob}{\ensuremath{\Omega_b}\xspace}
\newcommand{\Hzero}{\ensuremath{H_0}\xspace}
\newcommand{\ns}{\ensuremath{n_s}\xspace}
\newcommand{\sigeight}{\ensuremath{\sigma_8}\xspace}
\newcommand{\Seight}{\ensuremath{S_8}\xspace}
\newcommand{\wzero}{\ensuremath{w_0}\xspace}

\newcommand{\Aia}{\ensuremath{A_{\text{IA}}}\xspace}
\newcommand{\Mc}{\ensuremath{M_c}\xspace}

\newcommand{\xip}{\ensuremath{\xi_+}\xspace}
\newcommand{\xim}{\ensuremath{\xi_-}\xspace}
\newcommand{\xipm}{\ensuremath{\xi_\pm}\xspace}
\newcommand{\zetap}{\ensuremath{\zeta_+}\xspace}
\newcommand{\zetam}{\ensuremath{\zeta_-}\xspace}
\newcommand{\zetapm}{\ensuremath{\zeta_\pm}\xspace}
\newcommand{\Map}{\ensuremath{M_{ap}}\xspace}

\newcommand{\lcdm}{\ensuremath{\Lambda}CDM\xspace}
\newcommand{\wcdm}{\ensuremath{w}CDM\xspace}

\title{\texttt{SBi3PCF:} cosmology with Dark Energy Survey Year 3 cosmic shear}

\author[a]{David Gebauer \orcidlink{0009-0002-2889-3704},}
\author[b,c]{Anik Halder \orcidlink{0000-0002-0352-9351},}
\author[d]{Zhengyangguang Gong \mbox{(\begin{CJK}{UTF8}{gbsn}龚郑阳光\end{CJK}) \orcidlink{0009-0002-7361-4073}},}
\author[a]{Cora Uhlemann \orcidlink{0000-0001-7831-1579}}

\affiliation[a]{Fakultät für Physik, Universität Bielefeld, Postfach 100131, 33501 Bielefeld, Germany}
\affiliation[b]{Institute of Astronomy and Kavli Institute for Cosmology, University of Cambridge, Madingley Road, Cambridge CB3 0HA, UK}
\affiliation[c]{Jesus College, Jesus Lane, Cambridge, CB5 8BL, UK}
\affiliation[d]{Steward Observatory, University of Arizona, 933 North Cherry Avenue, Tucson, AZ 85721, USA}

\emailAdd{david.gebauer@uni-bielefeld.de}

\abstract{
We present the first cosmic shear data analysis of the integrated 3-point correlation function (i3PCF), a higher-order weak lensing statistic, and the first simulation-based inference (SBI) analysis of a three-point cosmic shear correlation function on survey data. Using \texttt{SBi3PCF}, our previously developed SBI framework for the i3PCF, we jointly constrain the $w$CDM parameter space with the cosmic shear 2-point correlation functions $\xi_\pm$ (2PCFs) and the i3PCFs $\zeta_\pm$ from the Dark Energy Survey Year 3 (DES Y3) weak lensing shape catalogue, from which we detect $\zeta_\pm$ with a signal-to-noise ratio of 12.8. Our forward model is built on the \texttt{CosmoGridV1} N-body simulations and includes the survey mask, shape noise, intrinsic alignment, baryonic feedback, photometric redshift uncertainty and shear calibration bias. The analysis was carried out blind, with all choices frozen, and only unblinded once coverage tests, tests of the accuracy of the learned likelihoods, and internal consistency tests had been passed. For the joint $\xi_\pm$ + $\zeta_\pm$ analysis we obtain $\Omega_m = 0.272^{+0.038}_{-0.033}$, $S_8 = 0.784^{+0.024}_{-0.021}$ and $w_0 = -1.04^{+0.25}_{-0.19}$. Adding $\zeta_\pm$ improves the figure of merit in the $\Omega_m$--$S_8$--$w_0$ parameter space by a factor of $1.59$ over $\xi_\pm$ alone, as predicted by simulated i3PCF analyses. This gain comes from third-order information on angular scales $\geq 15'$ and aperture radii $\geq 50'$, a regime where baryonic effects are subdominant. The constraints are consistent with the other DES Y3 analyses of the same catalogue and with independent lensing analyses, and differ by at most $2.4\sigma$ from CMB and BAO data (Planck 2018 + BAO, in $\Omega_m$--$S_8$--$w_0$). This shows that $\zeta_\pm$ can be analysed on cosmological survey data with a fully simulation-based approach, an important step towards Stage-IV surveys.
}

\begin{document}

\maketitle
\flushbottom

\FloatBarrier
\section{Introduction}

Cosmic shear, the coherent distortion imprinted in the observed shapes of background galaxies by the weak gravitational lensing (WL) effect of the foreground matter distribution \cite{BartelmannSchneider2001, Schneider2006, Kilbinger2015}, has become one of the most powerful probes of the large-scale structure (LSS) of the Universe. By analysing statistical correlations in the cosmic shear field, we can probe the distribution and the properties of the matter density field and constrain key cosmological parameters, such as the matter density \Om, the amplitude of matter density fluctuations \sigeight (often expressed through the structure growth parameter $\Seight = \sigeight \sqrt{\Om/0.3}$, and the dark energy equation of state parameter \wzero.

Current Stage-III WL surveys, such as the Dark Energy Survey (DES) \cite{DES_overview, DES_Y3_data, Gatti2021}, the Kilo-Degree Survey (KiDS) \cite{Kids_overview, Wright2024}, and the Hyper Suprime-Cam (HSC) survey \cite{HSC_overview}, have placed powerful constraints on these parameters, primarily by measuring two-point statistics (2-point correlation functions \xipm (2PCFs) and power spectra) of the shear field \cite{DESY3_results, Secco2022a, Amon2022b, Kids_results, HSC_results, Li2023, Anbajagane2025}. However, \xipm only probes the information contained in the second moments of the density field and is not sensitive to the higher-order moments of the non-Gaussian late-time matter distribution \cite{Bernardeau2002}. To extract this additional information, one has to turn to higher-order statistics (HOS), which have already been applied to weak lensing survey data, for example through third-order shear statistics measured on CFHTLenS \cite{Fu2014}, DES \cite{Secco2022b, Gomes2025_des_a, Gomes2025_des_b}, KiDS \cite{Burger2024, Porth2024, Linke2026} and HSC \cite{Sugiyama2025} data, moments and cumulative distribution functions of the DES Y3 mass maps \cite{Gatti2022, Anbajagane2023}, peak counts on DES \cite{Kacprzak2016, Zuercher2022}, KiDS \cite{Martinet2018, Harnois-Deraps2024} and HSC \cite{Marques2024} data, the convergence PDF on HSC \cite{Thiele2023, Novaes2025} data, persistent homology on DES \cite{Heydenreich2022} and the scattering transform on HSC data \cite{Cheng2025}. Beyond these, a range of further HOS is currently being developed, with the Euclid papers \cite{Euclid_HOWLS, Euclid_HOWLS_DR1} giving an overview, including the weak lensing convergence 1-point PDF \cite{Uhlemann2023, Giblin2023, Barthelemy2024, Castiblanco2024, Friedrich2025}, critical points of the convergence field \cite{Gong2025}, the parity-odd lensing trispectrum \cite{Chen2026} and machine-learning-derived summary statistics \cite{Gong2024, Lehman2024, Lehman2026}.

A practical way of accessing higher-order information is through the integrated 3-point correlation function (i3PCF) \cite{Halder2021, Halder2023, Gong2023}, which correlates \xipm measured locally in small patches of the sky with the aperture mass at the patch centres. It is mostly sensitive to squeezed triangle configurations of the lensing bispectrum and is measured directly from the cosmic shear field, without the need to reconstruct a convergence map, or to apply a full 3PCF estimator. In contrast to many other HOS, it can also be modelled analytically and interpreted physically using the response approach to perturbation theory \cite{Halder2022, Barreira_2017}.

Extracting the cosmological information contained in such HOS such as the i3PCF, however, requires a likelihood function, which in general is not known analytically. In simulation-based inference (SBI), the likelihood of the summary statistic is instead learned from a suite of forward-modelled simulations that include the survey and noise properties, and systematic effects using generative neural networks such as normalising flows \cite{Papamakarios2017, Alsing2019}. This avoids approximations such as the widely used Gaussian likelihood, which might not hold for HOS. SBI has been applied to cosmic shear data from DES SV \cite{Jeffrey2021} and KiDS-1000 \cite{Fluri2022, Wietersheim-Kramsta2025}, and in a number of DES Y3 analyses using wavelet harmonics, scattering transforms and moments of mass maps \cite{Gatti2024b, Gatti2025}, persistent homology \cite{Prat2025}, neural compression at the map level \cite{Thomsen2026, Jeffrey2025}, and combinations of the power spectrum with neural network summaries \cite{Williamson2026}.

In the first paper of this series \cite{SBi3PCF1} (hereafter Paper~1), we developed and validated \texttt{SBi3PCF}, a complete SBI framework for a joint analysis of the cosmic shear 2PCFs \xipm and the i3PCFs \zetapm. Its forward model, built on the \texttt{CosmoGridV1} N-body simulations \cite{Kacprzak2023}, generates realistic DES Y3-like cosmic shear mocks including the survey mask, shape noise, and a range of crucial systematic effects such as intrinsic alignments and baryonic feedback. There, we have shown on simulated measurements that the pipeline produces well-calibrated posteriors, recovers the true cosmological parameters without bias, and that adding \zetapm to a standard \xipm analysis significantly improves the cosmological constraining power.

In this work, we apply this framework to the DES Y3 data and present the first cosmological analysis of \zetapm in weak lensing survey data. We measure \xipm and \zetapm from the DES Y3 weak lensing shape catalogue \cite{Gatti2021} and infer posteriors in the \wcdm parameter space spanned by \texttt{CosmoGridV1}, for each statistic individually as well as for their joint analysis. To avoid confirmation bias, we carry out the analysis blind. This means that all analysis choices are frozen, and that a series of tests (coverage tests, tests of the accuracy of the learned likelihoods and internal consistency tests on the blinded measurement) has to be passed before we look at the cosmological constraints \cite{Muir2020}. We then present our constraints on \Om, \sigeight, \Seight and \wzero and compare them with the other DES Y3 analyses of the same source catalogue, the latest DES Y6 3$\times$2pt results \cite{DESY6_results}, and also with external data: KiDS-1000 \cite{Troester2021}, KiDS-Legacy \cite{Kids_results, Reischke2026}, Planck \cite{Planck2018} and DESI \cite{DESI_DR2}. Aside from the constraints themselves, this analysis prepares \zetapm for the beyond-two-point SBI analyses of the upcoming Stage-IV surveys Euclid \cite{Euclid_overview} and Rubin LSST \cite{LSST_overview}, where the statistical power of the data will require the treatment of systematic effects such as intrinsic alignments \cite{joachim_TATT_2025}, baryonic feedback \cite{Bigwood2024, Lucie-Smith_2025} and photometric redshift outliers \cite{Mill2025} in the forward model (for a broader overview of these systematic effects, see Section~2.2.1 of Ref.~\cite{CosmoVerse2025}).

We structure this paper as follows. In Section~\ref{sec:data}, we describe the DES Y3 data, the measurement of the correlation functions and the simulation-based forward model. Section~\ref{sec:sbi} briefly summarises our SBI pipeline, for which we refer the reader to Paper~1 for details. We present the pre-unblinding tests in Section~\ref{sec:validation}, our cosmological constraints in Section~\ref{sec:constraints} and the comparison with external data in Section~\ref{sec:comparison}, and conclude in Section~\ref{sec:conclusions}.

\FloatBarrier
\section{Data and simulations}\label{sec:data}

The forward model and the measurements are identical to Paper~1, so that we only repeat the essentials here. For further details, we refer the reader to Sections~2 and 3 of Paper~1 \cite{SBi3PCF1}.

\subsection{DES Y3}\label{sec:desy3}

The Dark Energy Survey (DES) spans $\sim$5000 deg$^2$ in the $grizY$ filters, observed with the Dark Energy Camera at the 4m Blanco telescope \cite{DES_overview}. This work uses data from the first three years of observations (Y3) with the original data products, i.e.~the Gold catalogue \cite{DES_Y3_data}, the weak lensing shape catalogue \cite{Gatti2021}, the PSF modelling \cite{Jarvis2021}, the image simulations used for the shear calibration \cite{MacCrann2022} and the SOMPZ redshift calibration \cite{Myles2021}. The only DES Y3 data product that we do not use is the shear-ratio (SR) likelihood \cite{Sanchez2022}, as we have neither a lens sample nor an analytic likelihood into which it could enter.

The DES Y3 weak lensing shape catalogue \cite{Gatti2021} uses \textsc{Metacalibration}-based \cite{Huff2017, Sheldon2017} shear estimates for 100,204,026 galaxies over an effective area of 4143 deg$^2$, with a weighted effective number density of 5.59 arcmin$^{-2}$ in the definition of Ref.~\cite{Heymans2012}. The galaxies are divided into the four DES Y3 tomographic source bins \cite{Myles2021}, with the per-bin number densities, mean redshifts and ellipticity dispersions of Table~I of Ref.~\cite{Amon2022b}, as in Paper~1. The redshift distribution of each bin was calibrated via SOMPZ \cite{Myles2021} and is shown in Figure~2 of Paper~1. The Gaussian priors on the per-bin shear calibration biases $m_i$ listed in Table~\ref{tab:prior_range} are taken from the image simulations of \cite{MacCrann2022}. They include a 2--3\% residual bias due to blending, which enters through the non-zero means of the priors and is marginalised over in the analysis.

We have pixelised the catalogue onto \texttt{HEALPIX} \cite{Gorski2005} maps with $\texttt{NSIDE} = 512$, corresponding to a pixel scale of $6.9'$ (pixel area of 47 arcmin$^2$). For each tomographic bin, we subtract the mean shear and apply the \textsc{Metacalibration} response, after which the value of a pixel is the weighted combination of the ellipticities of all galaxies falling within it, with the pixel weight given by the sum of the individual galaxy weights. Pixels outside the DES Y3 footprint are set to have zero weights, such that the mask is applied to the data and the mocks in the same way. Since every mock of the forward model carries the same per-pixel weights (Section~\ref{sec:forward_model}), the observed data enter the measurement in exactly the same pixelised form as the mocks.

We have blinded our analysis at the catalogue level, following the DES Y3 scheme \cite{Gatti2021}. All galaxy ellipticities were rescaled by an unknown multiplicative factor, and every subsequent step (the pixelisation, the measurement, the compression and the inference) was run on this blinded catalogue. We did not apply any additional parameter shifts to the posteriors \cite{Muir2020}. Instead, all axis values and numerical summaries of the blinded posteriors were hidden, and these posteriors were never shown together with literature, fiducial or other chains. This differs from the other DES Y3 SBI analyses, which blinded at the posterior level, either by not looking at the posteriors of the data \cite{Gatti2025, Prat2025} or by shifting them by an unknown offset \cite{Jeffrey2025, Williamson2026}, and is closest to the DES Y3 \xipm + $\langle M_{ap}^3 \rangle$ analysis \cite{Gomes2025_des_b}, which also measured its data vector on a catalogue with rescaled ellipticities. Because the blinding acts on the catalogue, the measured data vector enters every test of Section~\ref{sec:validation} unchanged. All analysis choices were frozen before these tests, and the unblinded catalogue was processed only after all of them had been passed. All figures and numbers of Sections~\ref{sec:data}--\ref{sec:validation} were first produced with the blinded measurement and replaced by the unblinded ones afterwards, and after unblinding no changes were made to the pipeline, the analysis choices, the scale cuts or the validation criteria.

\subsection{Correlation Functions}

We use the cosmic shear 2PCFs \xipm and the i3PCFs \zetapm, which correlate the local \xipm measured in small, roughly degree-scale sky patches with the aperture mass \Map at the patch centres \cite{Halder2021, Halder2023, Gong2023}. \xipm are estimated from the tangential and cross components of the pixel shears relative to the line connecting a pixel pair, as the weighted average over all pixel pairs within a separation bin, with the pixel weights of Section~\ref{sec:desy3}. \Map is estimated as the weighted sum of the tangential shear of all pixels around the patch centre with the compensated filter of Refs.~\cite{Crittenden2002, Jarvis2004, Schneider2005} and aperture radius $\theta_{\rm ap}$, and \zetapm as the mean product of the local \xipm and aperture masses over all patches, minus the product of their averages over the footprint. The estimators are given in Eqs.~(3.4)--(3.8) of Paper~1 and are applied in the same form to the pixelised DES Y3 maps and to every mock.

The measurement setup is identical to Paper~1. We measure the global \xipm in 8 logarithmically spaced bins in the angular separation $\alpha$ between $15'$ and $250'$. The lower limit is about twice the pixel scale at $\texttt{NSIDE} = 512$ and therefore excludes the small scales that would be most affected by baryonic feedback, while the upper limit matches the DES Y3 \xipm setup \cite{DESY3_results, Secco2022a}. For \zetapm we use five aperture radii $\theta_{\rm ap} \in \{50', 70', 90', 110', 130'\}$. The local \xipm are measured in the same binning as the global ones, truncated at $\alpha_{\max} \le 2\theta_{\rm ap} - 5'$, which leaves 5, 6, 6, 7 and 8 angular bins for the five apertures. The patch centres are placed on a \texttt{HEALPIX} grid at $\texttt{NSIDE} = 64$ for $\theta_{\rm ap} = 50'$ and $70'$ and at $\texttt{NSIDE} = 32$ for the larger apertures, of which we keep those centres for which both the patch (radius $\theta_{\rm ap}$) and the region over which the filter is evaluated (radius $5\theta_{\rm ap}$) are at most 50\% masked, which leaves between $5763$ ($50'$) and $846$ ($130'$) overlapping patches. The patch centres are identical for the DES Y3 measurement and for every mock footprint.

For both statistics we measure the auto- and cross-correlations of all tomographic bin combinations. For \xipm these are the 10 pairs $(i,j)$ with $i \le j$, and for \zetapm the 20 combinations $(i,j,k)$ with $i \le j \le k$, where \Map is measured in bin $i$ and the local \xipm between bins $j$ and $k$. The \xipm data vector thus has $160$ elements (10 tomographic bin combinations $\times$ 2 components $\times$ 8 angular bins), the \zetapm data vector 1280 (20 bin combinations $\times$ 2 components $\times$ 32 angular bins summed over the apertures), and the joint \xipm + \zetapm data vector a total length of 1440.

All measurements, on the DES Y3 data and on all mocks, are performed with the same pipeline as in Paper~1, i.e.~the global \xipm with \texttt{TreeCorr} \cite{Jarvis2004, Jarvis2019} on the pixelised maps and \zetapm with our GPU code \texttt{PixPoint} (formerly \texttt{CosmoFuse}; Appendix~A of Paper~1). As the same estimators are applied to identically pixelised maps, the measured data vector is directly comparable to mock data vectors produced by our forward model.

The measured \xipm and \zetapm are shown in Appendix~\ref{sec:uncompressed} together with the posterior predictive means of the analyses of Section~\ref{sec:constraints}. With the covariance estimated from the 16,000 fiducial footprints of the forward model (Section~\ref{sec:forward_model}), the signal-to-noise ratio, $\mathrm{S/N} = \sqrt{x^T C^{-1} x - N_{\rm d}}$, which removes the noise contribution of the $N_{\rm d}$ data points \cite{Secco2022b}, is 31.0 for \xipm, 12.8 for \zetapm and 33.0 for the joint data vector.

Finally, we do not perform any additional PSF or B-mode null tests on \zetapm, since these tests have already been carried out at the level of the DES Y3 shape catalogue \cite{Gatti2021, Jarvis2021, Amon2022b, Secco2022a}, which we use as released. \Map and the local \xipm are linear and quadratic, respectively, in the validated pixelised shear field as the global \xipm, so that PSF residuals and B-modes at the level allowed by the catalogue tests, well below the statistical uncertainty of the DES Y3 \xipm, are also expected to be negligible for the shape-noise-dominated \Map measurements. Hence, we do not measure the B-mode combinations, i.e.~\Map from $\gamma_\times$ and $\xi_\times$.

\subsection{Forward Model}\label{sec:forward_model}

The forward model was developed and validated in Paper~1, which is described in detail in its Section~3. It is based on the \texttt{CosmoGridV1} particle shells, from which we create weak lensing mocks with the DES Y3 mask, per-pixel shape noise from the observed catalogue and systematic effects added at the map level. In the following, we summarise the essential parts of the forward model for the data analysis and refer the reader to Paper~1 for the details.

The forward model varies 17 parameters: six \wcdm parameters, two baryonification \cite{Schneider2015, Schneider2019} parameters capturing baryonic feedback, one intrinsic alignment amplitude (in the NLA model \cite{HirataSeljak2004, BridleKing2007}) and four parameters each for the shear calibration bias and the mean photometric redshift shift.
Table~\ref{tab:prior_range} shows the prior ranges used in this work, which are the same as in Paper~1, together with the fiducial parameter values. For the cosmological, baryonification and intrinsic alignment parameters we use uniform priors. Apart from these, there are further restrictions in the \Om--\sigeight and \Om--\wzero planes, which are inherited from the \texttt{CosmoGridV1} grid (see Section~2 of Ref.~\cite{Kacprzak2023}). The restriction in the \Om--\sigeight plane was chosen to concentrate on the simulations along the known degeneracy direction, and the latter in the \Om--\wzero plane removes those phantom DE cosmologies ($\wzero < -1$) for which the relativistic treatment of the simulations breaks down before $z = 0$ (see Appendix~A of Ref.~\cite{Fluri2022}). On the shear calibration biases $m_i$ and the photometric redshift shifts $\Delta z_i$ we use the DES Y3 Gaussian priors following Refs.~\cite{DESY3_results, Secco2022a}. As in Paper~1, \Aia, $m_i$ and $\Delta z_i$ are drawn from their priors once per full-sky map and shared by its four footprints, with \Aia drawn from a Sobol sequence \cite{Sobol1967} and $m_i$ and $\Delta z_i$ from their Gaussian priors. The fiducial values of Table~\ref{tab:prior_range} are those of the \texttt{CosmoGridV1} fiducial cosmology, at which the covariance and the parameter derivatives entering the compression (Section~\ref{sec:sbi}) are evaluated. These priors are used for every inference in Section~\ref{sec:constraints}. For a full corner plot including a chain sampled from the prior alone, see Figure~\ref{fig:corner_full} in Appendix~\ref{sec:full_posterior}.
\begin{table}[H]
    \centering
    \begin{tabular}{l l l@{\ }l@{\,}r@{,\ }l@{\,}l}
        \hline
        \multicolumn{1}{c}{\textbf{Parameter}} & \multicolumn{1}{c}{\textbf{Fiducial}} & \multicolumn{5}{c}{\textbf{Prior}} \\
        \hline
        \Om & $\phantom{-}0.26$ & $\in$ & $[$ & $0.10$ & $\phantom{-}0.50$ & $]$ \\
        \Ob & $\phantom{-}0.0493$ & $\in$ & $[$ & $0.03$ & $\phantom{-}0.06$ & $]$ \\
        \Hzero & $\phantom{-}67.3$ & $\in$ & $[$ & $64.0$ & $\phantom{-}82.0$ & $]$ \\
        \ns & $\phantom{-}0.9649$ & $\in$ & $[$ & $0.87$ & $\phantom{-}1.07$ & $]$ \\
        \sigeight & $\phantom{-}0.84$ & $\in$ & $[$ & $0.40$ & $\phantom{-}1.40$ & $]$ \\
        \wzero & $-1.0$ & $\in$ & $[$ & $-2.00$ & $-0.33$ & $]$ \\
        \hline
        $\log M_c^0$ & $\phantom{-}13.82$ & $\in$ & $[$ & $12.0$ & $\phantom{-}15.0$ & $]$ \\
        $\nu$ & $\phantom{-}0.0$ & $\in$ & $[$ & $-2.00$ & $\phantom{-}2.00$ & $]$ \\
        \hline
        \Aia & $\phantom{-}0.0$ & $\in$ & $[$ & $-5.00$ & $\phantom{-}5.00$ & $]$ \\
        \hline
        $m_1$ & $-0.0063$ & $\mathcal{N}$ & $($ & $-0.0063$ & $\phantom{-}0.0091$ & $)$ \\
        $m_2$ & $-0.0198$ & $\mathcal{N}$ & $($ & $-0.0198$ & $\phantom{-}0.0078$ & $)$ \\
        $m_3$ & $-0.0241$ & $\mathcal{N}$ & $($ & $-0.0241$ & $\phantom{-}0.0076$ & $)$ \\
        $m_4$ & $-0.0369$ & $\mathcal{N}$ & $($ & $-0.0369$ & $\phantom{-}0.0076$ & $)$ \\
        $\Delta z_1$ & $\phantom{-}0.0$ & $\mathcal{N}$ & $($ & $0.0$ & $\phantom{-}0.018$ & $)$ \\
        $\Delta z_2$ & $\phantom{-}0.0$ & $\mathcal{N}$ & $($ & $0.0$ & $\phantom{-}0.015$ & $)$ \\
        $\Delta z_3$ & $\phantom{-}0.0$ & $\mathcal{N}$ & $($ & $0.0$ & $\phantom{-}0.011$ & $)$ \\
        $\Delta z_4$ & $\phantom{-}0.0$ & $\mathcal{N}$ & $($ & $0.0$ & $\phantom{-}0.017$ & $)$ \\
        \hline
    \end{tabular}
    \caption{The prior ranges used for inference and the fiducial parameter values. The priors for the cosmological, baryonification, and intrinsic alignment parameters are uniform flat priors, denoted by $\in [\,\min, \max\,]$. For the shear calibration and photometric redshift uncertainty parameters, we use Gaussian prior distributions, denoted by $\mathcal{N}(\,\text{mean}, \text{std}\,)$. \Hzero is given in km\,s$^{-1}$\,Mpc$^{-1}$.}
    \label{tab:prior_range}
\end{table}

The \texttt{CosmoGridV1} \cite{Kacprzak2023} suite of N-body simulations was run with \texttt{PKDGRAV3} \cite{Potter2017} and is provided as full-sky \texttt{HEALPIX} particle shells at $\texttt{NSIDE} = 512$, with a mass resolution that is sufficient for the statistics and scales ($\geq 15'$) considered here (see the discussion in Section~3.1 of Paper~1) and three degenerate neutrino species with $\sum m_\nu = 0.06$\,eV. Throughout this work, we use the set of shells in which baryonic feedback has been included at the map level via the shell baryonification method \cite{Schneider2015, Schneider2019, Kacprzak2023} with the $(\Mc, \nu)$ of each grid cosmology. Here, \Mc is the characteristic halo mass below which feedback has pushed a large fraction of the gas out of the haloes, flattening their gas profiles \cite{Schneider2019}, on which we place a log-uniform prior, and $\nu$ describes its redshift evolution as $M_c(z) = M_c^0 (1+z)^\nu$ \cite{Fluri2022, Kacprzak2023}. The simulation set consists of a grid of 2500 flat \wcdm cosmologies with 7 realisations each, made up of two Sobol-sampled \cite{Sobol1967} sets of 1250 cosmologies that span the full prior box of Table~\ref{tab:prior_range} and a narrower central region, and of a fiducial set of 200 realisations at the fiducial cosmology and at a $\pm$ step in each of the 6 cosmological and 2 baryonification parameters. The test cosmologies of Section~\ref{sec:sbi}, which are excluded from training, are taken from the full-box set.

From the particle shells, we create the weak lensing maps with \texttt{UFalcon2} \cite{Sgier2019, Sgier2021, Reeves2024} as described in Section~3.2 of Paper~1: the shells are weighted by the lensing kernel of each tomographic bin in the Born approximation, the resulting convergence maps are converted to shear maps through a full-sky spin-2 harmonic transform, akin to Kaiser-Squires reconstruction \cite{KaiserSquires1993}, and the DES Y3 mask is applied. By rotating each full-sky map, four non-overlapping footprints are extracted, leading to 2500 cosmologies $\times$ 7 realisations $\times$ 4 footprints $=$ 70,000 independent mocks for training. From the fiducial maps, $200$ realisations $\times$ 20 random orientations $\times$ 4 footprints $=$ 16,000 fiducial footprints are obtained, from which the covariance entering the compression (Section~\ref{sec:sbi}) is estimated. To each footprint, we then add per-pixel shape noise, obtained by randomly rotating the ellipticities of the galaxies in the observed DES Y3 catalogue and summing them up within each pixel, so that every mock carries the same per-pixel weights as the data, as in the other DES Y3 map-based SBI analyses \cite{Gatti2024b, Jeffrey2025, Thomsen2026}. The systematic effects are added at the map level as described in Section~2.4 of Paper~1: intrinsic alignments with the nonlinear alignment (NLA) model \cite{HirataSeljak2004, BridleKing2007} with a free amplitude \Aia, without redshift evolution ($\eta = 0$) and with $C_1 = 5 \times 10^{-14} h^{-2} M_\odot^{-1} {\rm Mpc}^3$ \cite{HirataSeljak2004, Joachimi2011}, i.e.~in the same normalisation as the DES Y3 NLA amplitude $a_1$ at the pivot redshift, the photometric redshift uncertainty as a coherent shift $n_i(z) \to n_i(z + \Delta z_i)$ of each bin's SOMPZ distribution when projecting the shells, and the shear calibration bias as a multiplicative factor $(1 + m_i)$ per bin on the shear maps. Reduced shear and source galaxy clustering are not part of the forward model, as they were shown to change the data vector by far less than the statistical uncertainty in Section~5.2 of Paper~1.

\FloatBarrier
\section{Simulation-based inference}\label{sec:sbi}

We infer the cosmological parameters by learning the likelihood of the compressed data vector from the forward-modelled mocks, i.e.~by neural likelihood estimation (NLE) \cite{Papamakarios2019, Alsing2019}, as developed and validated in Paper~1. The pipeline consists of three steps, which are described in detail in Section~4 of Paper~1: (i) a linear \texttt{MOPED} \cite{Heavens2000, Heavens2017} compression of the 1440-element data vector to 17 coefficients (Section~\ref{sec:compression}), (ii) a normalising flow trained on the parameters and compressed data vectors of the training set to estimate the likelihood $p(x \,|\, \theta)$ (Section~\ref{sec:nde}), and (iii) MCMC sampling of the posterior $p(\theta \,|\, x_{\rm obs}) \propto p(x_{\rm obs} \,|\, \theta)\, \pi(\theta)$ with the flat prior $\pi(\theta)$ of Table~\ref{tab:prior_range}, for which we use our own GPU implementation of the affine-invariant sampler \cite{Goodman2010} (Section~\ref{sec:sampling}). The SBI  pipeline does not assume a Gaussian likelihood or an analytical model of the data vector. The only place where the fiducial covariance is used is in the construction of the compression vectors (Section~\ref{sec:compression}).

Compared to Paper~1, the pipeline is now trained and tuned separately for every analysis variation, i.e.~\xipm, \zetapm, their combination and the data-vector variations of Section~\ref{sec:validation}, and then applied to the measured (blinded) DES Y3 data vector. We further use neural spline flows (NSFs) instead of masked autoregressive flows (MAFs), a different hyperparameter search space (see Appendix~\ref{sec:hyperparameters}), the binomial instead of the bootstrap uncertainty band for the coverage test (see Appendix~\ref{sec:coverage_tests}), and a faster MCMC sampler that runs many chains at once on a GPU. On a single NVIDIA A100 GPU, one trial of the hyperparameter search (Section~\ref{sec:nde}) takes about 10 minutes, so that its 300 trials amount to 2--3 GPU days per analysis variation. With the batched sampler, the 1,250 test and 800 fiducial posteriors of an analysis variation are sampled in about 3 and 2 hours, respectively, and the 10 million-sample DES Y3 chain in about 25 minutes.

\subsection{MOPED Compression}\label{sec:compression}

To reduce the dimensionality of our data vectors, we employ the \texttt{MOPED} compression \cite{Heavens2000, Heavens2017}, a linear transformation $y_\alpha = b_\alpha^T x$ that maximises the Fisher information of the data vector $x$ at the fiducial cosmology. It reduces the 1440-element joint data vector (160 for \xipm alone, 1280 for \zetapm alone) to 17 coefficients, one per forward-model parameter: the 6 cosmological parameters, \Mc, $\nu$, \Aia, the four $m_i$ and the four $\Delta z_i$. The compression vectors $b_\alpha \propto C^{-1} \partial \mu / \partial \theta_\alpha$, where $\mu(\theta)$ is the mean data vector at the parameters $\theta$, are Gram-Schmidt orthogonalised in the order of Table~\ref{tab:prior_range}, so that the coefficients are uncorrelated at the fiducial cosmology (Eqs.~4.4--4.5 of Paper~1). The covariance $C$ is estimated from the 16,000 fiducial footprints (Section~\ref{sec:forward_model}) and its inverse carries the Hartlap correction factor \cite{Hartlap2007} (for 16,000 footprints and 1440 elements the correction factor is 0.91). While this constant factor is not relevant for the \texttt{MOPED} coefficients, we still include it for the later goodness-of-fit tests in Section \ref{sec:validation}. The parameter derivatives $\partial \mu / \partial \theta_\alpha$ with respect to the cosmological and baryonification parameters come from the dedicated \texttt{CosmoGridV1} derivative simulations around the fiducial cosmology, as central finite differences of the noiseless mean over the 200 realisations of the $\pm$ simulations, and those with respect to \Aia, $m_i$ and $\Delta z_i$ from the fiducial maps reprocessed with the respective parameter shifted.\footnote{The step sizes are $\pm 0.01$ in \Om, $0.001$ in \Ob, $2$ in \Hzero, $0.02$ in \ns, $0.015$ in \sigeight, $0.05$ in \wzero, $0.1$ in $\log \Mc$, $\nu$, \Aia and $m_i$, and $0.01$ in $\Delta z_i$.} We compute a separate set of compression vectors for each analysis variation defined in Section~\ref{sec:validation}. For the variations that remove a tomographic bin, the derivative with respect to its shear calibration bias vanishes, so that the compressed data vector has 16 instead of 17 coefficients. 

We note that the compression is performed at the fiducial parameter set. It is lossless only for a Gaussian likelihood with a parameter-independent covariance and a data vector that depends linearly on the parameters. Away from the fiducial parameters, some information is lost, but no bias is introduced, since the likelihood of the compressed data vector is learned directly (see Paper~1).

\subsection{Neural Density Estimation}\label{sec:nde}

To learn the likelihood $p(x \,|\, \theta)$ of the \texttt{MOPED} coefficients, we use neural spline flows (NSFs) \cite{Durkan2019}, autoregressive flows whose transformations are monotonic rational-quadratic splines. They replace the masked autoregressive flows of Paper~1 and are implemented in \texttt{zuko} \cite{Rozet2024}, which we use through the \texttt{sbi} package \cite{sbi2020}. Both the parameters and the compressed data vectors are standardised with the mean and standard deviation of the simulation set before training. The base distribution of the flow is a standard normal, and the loss function is the negative mean log-likelihood of the training batch. As in Paper~1, we only provide the 9 parameters of interest, \Om, \Ob, \Hzero, \ns, \sigeight, \wzero, \Aia, \Mc and $\nu$, as labels. The remaining 8 systematic parameters ($m_i$ and $\Delta z_i$) are thereby implicitly marginalised over while learning the likelihood \cite{Alsing2019b}.

Of the 70,000 forward-modelled mocks, 60,000 are used for training, 7,500 for validation, and 2,500 are excluded from training (one footprint of one realisation per cosmology) as test set. The 1,250 of these belonging to the full-box Sobol set (Section~\ref{sec:forward_model}) form the test set of Section~\ref{sec:validation}. We note that training on a finite number of simulations adds noise to the learned likelihood, similar to the Dodelson-Schneider effect \cite{Dodelson2013} for covariance matrices estimated from simulations. Unlike for a Gaussian likelihood, however, this additional uncertainty propagates into the width of the posteriors \cite{Homer2025}.

The hyperparameters of the flows are tuned with \texttt{optuna} \cite{optuna_2019} using a \texttt{TPESampler} \cite{bergstra2011} for 300 trials, separately for every analysis variation, and we select the model with the highest likelihood on the validation set, i.e.~the lowest ``validation loss''. The search ranges, the fixed training settings and the selected hyperparameters of every analysis variation are listed in Appendix~\ref{sec:hyperparameters}. We use a single flow per analysis variation rather than an ensemble, as this already yields well-calibrated posteriors in the coverage tests (Appendix~\ref{sec:coverage_tests}). As these test the calibration on average over the test set, we have also applied the five next-best trials of the joint analysis to the unblinded DES Y3 data vector. In units of the 68\% half-width of the respective joint posterior, the posterior medians shift by at most $0.26$ in \Om, $0.16$ in \Seight and $0.14$ in \wzero, which is small compared to the statistical uncertainty.

\subsection{Sampling posteriors}\label{sec:sampling}

We use a flat prior over the parameter space spanned by \texttt{CosmoGridV1} (Table~\ref{tab:prior_range}), including the restrictions in the \Om--\sigeight and \Om--\wzero planes, which are enforced as a hard boundary with a log-probability of $-\infty$ outside the prior. The posteriors are then sampled by MCMC with the learned likelihood. For this, we have written a dedicated GPU implementation of the affine-invariant ensemble sampler with the stretch move ($a = 2$) \cite{Goodman2010} in \texttt{PyTorch} \cite{Paszke2019}, which evaluates the likelihood of all walkers at once and runs up to 200 chains simultaneously on a single GPU. This replaces the \texttt{emcee} \cite{Foreman2013} sampling of Paper~1 and makes the thousands of posteriors needed for the tests of Section~\ref{sec:validation} computationally feasible.

Each chain has 1,000 walkers, which are initialised in a Gaussian ball of $0.3$ prior standard deviations around the \texttt{CosmoGridV1} fiducial parameters for the DES Y3 measurement, or around the true parameters for the simulated measurements. For the simulated measurements (test and fiducial), each walker runs for 5,000 steps, of which the first 2,000 are discarded as burn-in and the remainder is thinned by 3, giving $10^6$ samples per posterior. For the DES Y3 measurement, each walker runs for 70,000 steps with 20,000 steps of burn-in and a thinning of 5, giving $10^7$ samples per posterior. With an integrated autocorrelation time of $\tau \approx 390$ steps per walker for the DES Y3 joint chain, this corresponds to about $1.3 \times 10^5$ independent samples. The simulated-measurement chains retain $\approx 8\tau$ per walker, which is sufficient for the medians and 68\% intervals used in the validation tests.

From the chains, we derive the parameter $\Seight = \sigeight \sqrt{\Om/0.3}$ and quote posterior medians with 68\% credible intervals. In addition, we determine the maximum of the posterior (MAP), i.e.~the best-fitting parameters, of every chain by maximising the log-posterior, the flow log-likelihood plus the prior, with the \texttt{Adam} optimiser \cite{Kingma2017} in \texttt{PyTorch}: starting from the 16 chain samples with the highest posterior probability, we take 300 steps with a learning rate of $10^{-2}$ and keep the best end point. The MAP is used for the goodness-of-fit tests of Section~\ref{sec:validation} and quoted in Section~\ref{sec:constraints}. The same pipeline, applied to the test simulations, is used for the pre-unblinding tests described in the next section.

\FloatBarrier
\section{Pre-unblinding tests}\label{sec:validation}

Before unblinding the cosmological constraints, the analysis had to pass a two--stage series of tests. First, the statistical calibration of the posteriors and the accuracy of the learned likelihoods were tested on simulated measurements. Second, the internal consistency between analysis variations, their goodness of fit, and posterior predictive distribution (PPD) tests \cite{Doux2021} were evaluated on the blinded measurement. The analysis variations are the three main analyses, i.e.~\xipm, \zetapm and their combination, and five variations of the joint \xipm + \zetapm data vector, in which either one tomographic bin (`no BIN1' to `no BIN4') or the two smallest angular bins (`no small scales') are removed. All criteria were fixed before assessing the blinded results, and every test was passed before unblinding. In this section, we briefly summarise the tests and their outcome. Their full description, together with all figures and tables, is given in Appendix~\ref{sec:app_validation}, where we show the unblinded measurement and quote the corresponding blinded results in the text. After unblinding none of the test outcomes changed. The unblinding criteria, in the order in which they were checked, and their results are:
\begin{enumerate}[label=(\roman*)]
    \item The coverage of all eight inference pipelines tested with \texttt{TARP} \cite{Lemos2023} lies within the $2\sigma$ binomial band (Appendix~\ref{sec:coverage_tests}) with all pipelines passing the coverage tests. The largest deviation of the empirical coverage from the credibility level $\alpha$ is $0.007$, compared to a $2\sigma$ band of $0.028$ for which $\alpha = 0.5$.
    \item The learned likelihoods are consistent with the $\chi^2$ expectation on the test set, verified by computing the squared Mahalanobis distance of each test data vector with respect to its own learned likelihood (this is a qualitative criterion, Appendix~\ref{sec:app_likelihoods}). While the test shows that the learned likelihoods are neither systematically too wide nor too narrow, the exaggerated tail reveals a (small) number of poorly fit cosmologies at the corners of the prior. The DES Y3 measurement itself lies at the 56th (joint), 60th (\xipm) and 73rd (\zetapm) percentile of $\chi^2$, i.e.~the learned likelihoods describe the data measurement to be indistinguishable from a typical mock.
    \item The blinded compressed data vector lies within the range spanned by the simulations and its log-posterior at the MAP falls inside the test-set distribution (qualitative, Appendix~\ref{sec:app_loglike}). The measurement lies at the 28th (joint), 20th (\xipm) and 24th (\zetapm) percentile of the test set, and the \texttt{MOPED} coefficients for \Om, \sigeight and \wzero deviate from the mean of the learned likelihood at the MAP by at most $0.13\sigma$. The only coefficients deviating by more than $2\sigma$ are the redshift evolution $\nu$ of the baryonification ($+2.9\sigma$) and the shear calibration bias $m_3$ ($-2.1\sigma$).
    \item The posteriors of the blinded measurement are consistent between the data-vector variations (qualitative, without a numerical threshold, Appendix~\ref{sec:app_consistency}). The largest shift of a posterior median relative to the joint analysis is the upward \Seight shift of `no BIN3' ($+2.2\sigma$), where $\sigma$ denotes the 68\% half-width of the joint posterior, followed by `no small scales' ($+1.3\sigma$). All other shifts in the cosmological parameters are at most $1.1\sigma$.
   \item The calibrated PPD $p$-value of the test statistic $T^{\rm cos}$ on the cosmological \texttt{MOPED} coefficients, i.e.~those of \Om, \sigeight and \wzero, satisfies $\tilde{p}(T^{\rm cos}) > 0.01$ for every PPD test, following the criterion used by DES \cite{Doux2021} (Appendix~\ref{sec:app_ppd}). We find $\tilde{p}(T^{\rm cos}) = 0.08$--$0.89$ for the five data-vector variations, the \xipm-only and \zetapm-only analyses and the two cross-prediction tests between \xipm and \zetapm, with the lowest value again for `no BIN3'. The parameters inferred from any subpart of the data vector therefore correctly predict the removed part of the measurement in \Om, \sigeight and \wzero.
\end{enumerate}

Two of these results are relevant for the following sections. First, the same PPD statistic evaluated on the full compressed data vector (i.e.~of all 17 MOPED coefficients), $\tilde{p}(T)$, is more marginal, with $0.021$ for the goodness of fit of the joint analysis and $0.001$ for `no BIN3'. This tension is concentrated in the $\nu$ and $m_3$ coefficients named above, while the cosmological coefficients show no tension. Our consistency criterion excludes these two directions, since the learned likelihood is not reliable in them: the parameter value of $m_3$ is never passed as input to the flow during training, and the data vector is barely sensitive to $\nu$, so that the likelihood in this direction is prior-dominated (Appendix~\ref{sec:app_ppd}). Second, the upward \Seight shift of `no BIN3', the $-2.1\sigma$ deviation of $m_3$ and the lowest $\tilde{p}(T^{\rm cos})$ all indicate that the amplitude of the third tomographic bin is slightly lower than what the joint model predicts, which pulls \Seight down when this bin is included. Similarly, removing the two smallest angular scales, $15'$ and $30'$, raises \Seight by $1.3\sigma$ and lowers \Mc by $0.7\sigma$, meaning these scales prefer a lower amplitude and stronger baryonic feedback. The same effect can be seen on the fiducial simulations, where the baryonic model is correct by construction and the shifts in \Seight and \Mc are anti-correlated. Both of these shifts in \Seight and \wzero move along the \Seight--\wzero degeneracy seen in Figure~\ref{fig:corner}.

\FloatBarrier
\section{Cosmological constraints}\label{sec:constraints}

In this section, we present the unblinded DES Y3 constraints of our three analyses on the cosmological parameters (Section~\ref{sec:constraints_cosmo}) and on the systematic parameters (Section~\ref{sec:constraints_systematics}), and quantify the improvement of the joint analysis over \xipm alone (Section~\ref{sec:constraints_improvement}).

\subsection{Constraints on \texorpdfstring{\Om, \Seight and \wzero}{Omega\_m, S8 and w0}}\label{sec:constraints_cosmo}

\begin{figure}[H]
    \centering
    \includegraphics[width=0.8 \linewidth]{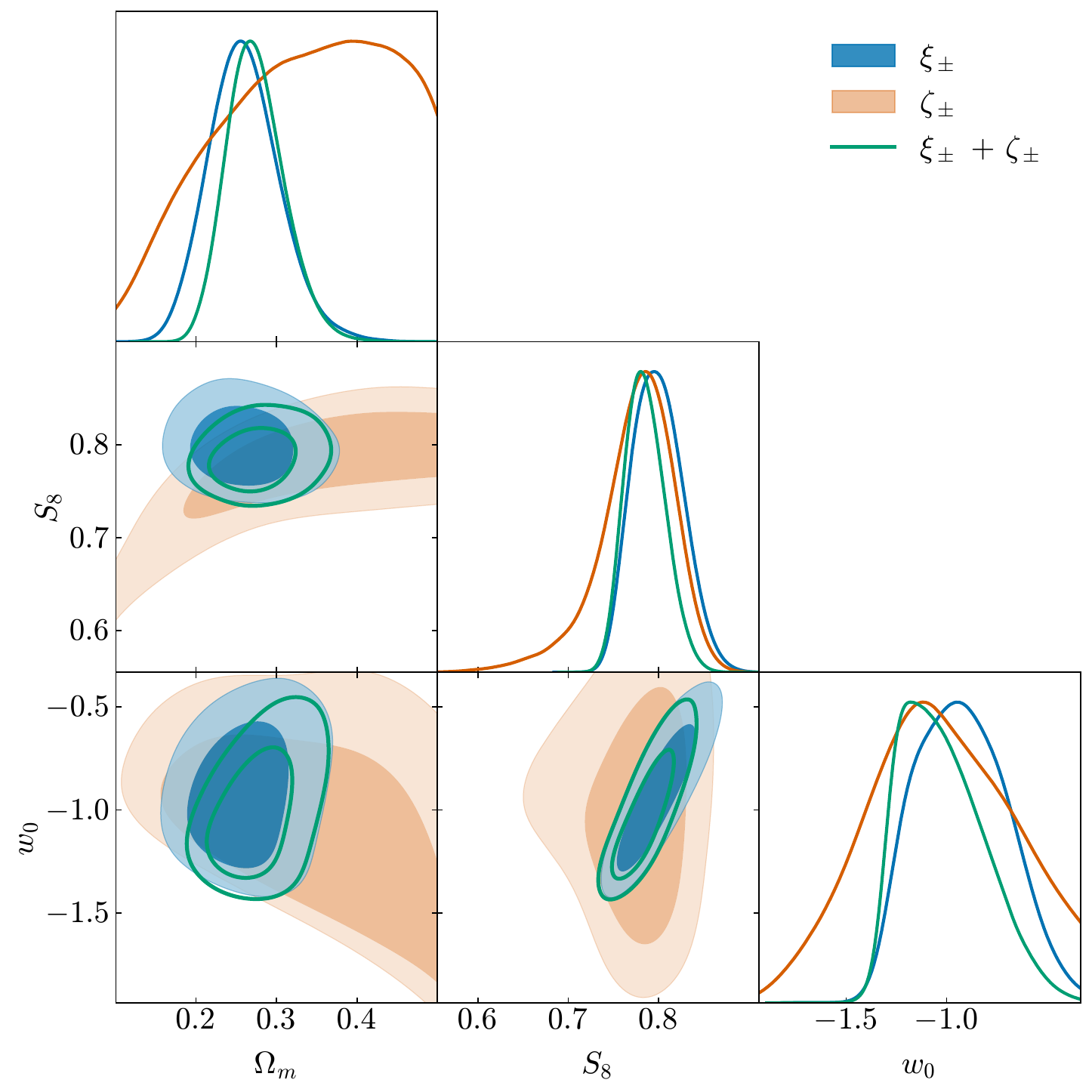}
    \caption{Marginalised posterior distributions of \Om, \Seight and \wzero from the DES Y3 measurement. The \xipm analysis is shown in blue, the \zetapm analysis in orange, and the joint \xipm + \zetapm analysis as green unfilled contours. The 1D marginalised posteriors are shown on the diagonal, and the 2D joint posteriors are shown as 68\% and 95\% credible contours.}
    \label{fig:corner}
\end{figure}

Figure~\ref{fig:corner} shows the marginalised DES Y3 posteriors of \Om, \Seight and \wzero for our three main analyses, and Table~\ref{tab:constraints} lists the corresponding medians and 68\% credible intervals of \Om, \Seight, \sigeight and \wzero. The full posterior of all sampled parameters is shown in Appendix~\ref{sec:full_posterior}. The joint \xipm + \zetapm analysis clearly tightens the constraints compared to \xipm alone, most of all for \sigeight and in the \Om--\Seight plane. The \zetapm-only posterior is broader and fully consistent with the other two analyses though its degeneracy direction in the \Om--\Seight plane differs from that of the \xipm posterior. \xipm constrains \Seight itself, while \zetapm constrains another combination of \sigeight and \Om ($\sigeight \Om^{\alpha}$ with $\alpha > 0.5$). This is in line with theoretical expectations, as at leading order \zetapm is proportional to $\sigeight^4$ on large scales, while \xipm is sensitive to $\sigeight^2$ \cite{Bernardeau2002, Halder2021}, and our findings in previous simulated analyses (Section~5.4 of Paper~1). The joint posterior lies at the intersection of the two, which is where the improvement in \Om and in the figure of merit come from. Adding \zetapm narrows and shifts the contours, as listed in Table~\ref{tab:constraints}: compared to \xipm alone, \Seight decreases by $0.6\sigma$, \wzero by $0.5\sigma$, and \Om increases by $0.4\sigma$.

\begin{table}[H]
    \centering
    \begin{tabular}{|l|cccc|c|}
        \hline
        \textbf{Analysis} & \Om & \Seight & \sigeight & \wzero & $\chi^2_{\rm MAP} / N_{\rm d}$ \\
        \hline
        \xipm ($N_{\rm d} = 160$)           & $0.259^{+0.045}_{-0.042}$ & $0.798^{+0.029}_{-0.027}$ & $0.861^{+0.089}_{-0.074}$ & $-0.93^{+0.25}_{-0.23}$ & $1.11$ (54th) \\[2pt]
        \zetapm ($N_{\rm d} = 1280$)         & $0.341^{+0.105}_{-0.121}$ & $0.780^{+0.035}_{-0.043}$ & $0.740^{+0.138}_{-0.087}$ & $-1.05^{+0.38}_{-0.35}$ & $1.10$ (70th) \\[2pt]
        \xipm + \zetapm ($N_{\rm d} = 1440$) & $0.272^{+0.038}_{-0.033}$ & $0.784^{+0.024}_{-0.021}$ & $0.823^{+0.058}_{-0.052}$ & $-1.04^{+0.25}_{-0.19}$ & $1.11$ (69th) \\
        \hline
    \end{tabular}
    \caption{Marginalised constraints (median and 16th--84th percentiles) on \Om, \Seight, \sigeight and \wzero from the DES Y3 measurement. The last column gives the goodness of fit in data space, i.e.~the reduced $\chi^2$ of the measured data vector of $N_{\rm d}$ elements with respect to the posterior predictive mean of the forward model at the MAP, and, in parentheses, its percentile among the same statistic for the 1250 test simulations (see Appendix~\ref{sec:uncompressed}).}
    \label{tab:constraints}
\end{table}

The best-fitting parameters, i.e.~the MAP in the 9-dimensional parameter space obtained by maximising the log-posterior (Section~\ref{sec:sbi}), at which the goodness-of-fit tests in Section~\ref{sec:validation} evaluate the model, are listed in Table~\ref{tab:map} in Appendix~\ref{sec:full_posterior}. The MAP is a point estimate in the full parameter space and not a marginalised quantity, and the prior-dominated parameters lie at the prior boundaries, so that the MAP in these directions is not determined by the data. For the joint analysis, the MAP of \wzero lies $0.6\sigma$ below the median, which illustrates the projection effects we expect for posteriors that lie along the curved \Om--\wzero degeneracy and are cut by the prior boundary: the medians in Table~\ref{tab:constraints} are affected by the volume of the posterior, while the MAP is not. 

\subsection{Astrophysical systematic parameters}\label{sec:constraints_systematics}

For the intrinsic alignment amplitude, we find $\Aia = 2.7^{+1.4}_{-1.9}$ for the joint analysis, consistent with the \xipm analysis, while it is unconstrained for \zetapm alone. This is larger than the DES Y3 cosmic shear result \cite{Secco2022a} and than the values of the map-based SBI analyses \cite{Gatti2025, Jeffrey2025, Prat2025}, which lie between $0.2$ and $0.6$ (see Section~\ref{sec:comparison}). In our case, the constraint on \Aia is driven by the amplitude of the low-redshift bins of \xipm (the `no BIN1' variation loses the constraint entirely, see Table~\ref{tab:ablation_constraints}). The high \Aia has only a small effect on the cosmological parameters: \Aia and \Seight are uncorrelated in the joint posterior ($r = 0.15$), and removing the first tomographic bin loses almost all constraining power on \Aia while only slightly raising \Seight, with a scatter that is comparable to leaving out any of the other tomographic bins (Table~\ref{tab:ablation_shifts}). A large \Aia could also point to a strongly aligned galaxy population within the DES Y3 source sample, which a colour/galaxy-property-split analysis \cite{McCullough2026, Halder2026} could test. For baryonic feedback, we find $\log \Mc = 13.9^{+0.8}_{-1.1}$, with the median in the upper half of the prior, i.e.~only a mild preference for the strong-feedback half of the prior (Figure~\ref{fig:corner_full}), while $\nu$ remains unconstrained. This is expected, since the scales $\geq 15'$ contain little information on baryonic feedback.

\subsection{Improvement from adding \texorpdfstring{\zetapm}{zeta}}\label{sec:constraints_improvement}

Table~\ref{tab:improvement} quantifies the improvement of the joint analysis over \xipm alone for the DES Y3 data vector, for the individual parameters and for the figure of merit (FoM). As in Paper~1, the improvement for an individual parameter is largest in \sigeight and smallest in \wzero. This improvement for the DES Y3 data is typical: the FoM improvements lie between the 65th and the 77th percentile of the same quantity for the 800 fiducial noise realisations and for the 1250 test cases. We note that it is obtained on comparatively large angular scales, as the global and local \xipm start at $15'$ and the smallest aperture radius is $50'$, on which the forward model only weakly depends on baryonic feedback (see Section~\ref{sec:comparison}).
\begin{table}[H]
    \centering
    \setlength{\tabcolsep}{4pt}
    \begin{tabular}{|l|cccc|ccc|}
        \hline
        & \Om & \Seight & \sigeight & \wzero & FoM$(\Om, \Seight)$ & FoM$(\Om, \Seight, \wzero)$ & FoM$(\Om, \sigeight, \wzero)$ \\
        \hline
        Improvement & 22\% & 26\% & 49\% & 10\% & 54\% & 59\% & 99\% \\
        \hline
    \end{tabular}
    \caption{Improvement in constraining power of the joint \xipm + \zetapm analysis over \xipm alone for the DES Y3 measurement, in per cent. The improvement is defined as $r - 1$, where $r$ is the ratio of the widths of the 68\% credible intervals (\xipm over joint) for the individual parameters and the ratio of the figure of merit $\text{FoM} = 1/\sqrt{\det \text{Cov}}$ (joint over \xipm), which are larger than one when the joint analysis is tighter. We give the FoM in the \Om--\Seight plane, following the other DES Y3 higher-order analyses \cite{Gatti2025, Prat2025, Thomsen2026}, and in the two three-dimensional parameter spaces of Paper~1.}
    \label{tab:improvement}
\end{table}

We note that all factors in Table~\ref{tab:improvement} are ratios for the single DES Y3 data realisation. The median improvements over the simulated sets are a different quantity and consistent with Paper~1 \cite{SBi3PCF1}, considering that retraining and retuning for every analysis variation (Section~\ref{sec:sbi}) leads to minor changes. We also obtain the precision on the DES Y3 data that we expected from Paper~1: the widths of the constraints differ by less than 20\% from those of the simulated inference on a fiducial mock (Section~5.4 of Paper~1).\footnote{The median \Om--\sigeight--\wzero FoM improvement over the fiducial realisations is $78\%$ here compared to $64\%$ in Paper~1, the median \sigeight interval is $29\%$ narrower compared to $31\%$, and the \Seight interval of the joint analysis is $\pm 0.022$ here compared to $\pm 0.026$ there.} This agreement between the data and the simulated expectation of Paper~1, together with the analytical understanding of the i3PCF and its parameter dependence \cite{Halder2021, Halder2022, Gong2023}, gives us physical confidence in the SBI results beyond the statistical tests alone. 

\FloatBarrier
\section{Comparison with external data}\label{sec:comparison}

\begin{figure}[H]
    \centering
    \includegraphics[width=\linewidth]{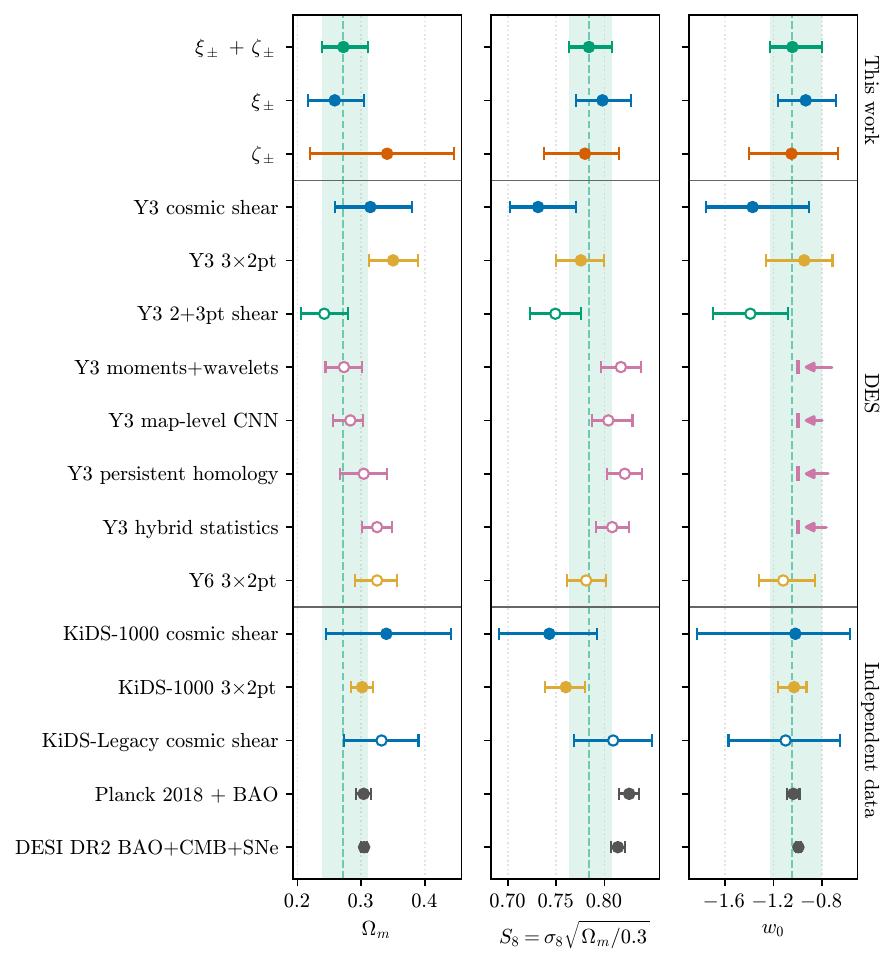}
    \caption{Marginalised constraints on \Om, \Seight and \wzero from published \wcdm analyses compared with the three analyses of this work. Filled markers show the medians with 16th--84th percentiles, which we have recomputed from the public chains, and open markers the published values (marginal mean and 68\% interval) for results without a public chain. Arrows pointing to the left in the \wzero panel indicate published 68\% upper limits, with the perpendicular line marking the non-phantom prior boundary $\wzero = -1$ of the respective analysis. The colours indicate the type of statistic: cosmic shear 2-point statistics are shown in blue like our \xipm, second- plus third-order shear statistics in green like our joint analysis, 3$\times$2pt in yellow, map-based SBI in pink, and the CMB and BAO results (Planck + BAO and DESI) in grey.}
    \label{fig:literature_comparison}
\end{figure}

Figure~\ref{fig:literature_comparison} compares the joint, the \xipm-only and the \zetapm-only constraints with published \wcdm results from the same source catalogue and from independent data. The constraints of every row are listed in Table~\ref{tab:literature_values}, recomputed from the public chains where available and taken from the publications otherwise. Table~\ref{tab:literature_setups} summarises the data, the likelihood, the treatment of intrinsic alignments and baryons, the amplitude and neutrino priors and the \wzero prior of every row, so that the differences in the priors and models between the analyses can be read from there.

\subsection{Published results}\label{sec:comparison_sample}

\begin{table}[H]
    \centering
    \begin{tabular}{|l|ccc|}
        \hline
        \textbf{Analysis} & \Om & \Seight & \wzero \\
        \hline
        \xipm + \zetapm (this work) & $0.272^{+0.038}_{-0.033}$ & $0.784^{+0.024}_{-0.021}$ & $-1.04^{+0.25}_{-0.19}$ \\[2pt]
        \hline
        DES Y3 cosmic shear \cite{Amon2022b, Secco2022a} & $0.314^{+0.066}_{-0.056}$ & $0.731^{+0.039}_{-0.029}$ & $-1.37^{+0.46}_{-0.39}$ \\[2pt]
        DES Y3 3$\times$2pt \cite{DESY3_results} & $0.350^{+0.039}_{-0.038}$ & $0.776^{+0.024}_{-0.026}$ & $-0.95^{+0.23}_{-0.31}$ \\[2pt]
        DES Y3 2+3pt shear$^*$ \cite{Gomes2025_des_b} & $0.242^{+0.038}_{-0.037}$ & $0.749^{+0.027}_{-0.026}$ & $-1.39 \pm 0.31$ \\[2pt]
        DES Y3 moments+wavelets$^*$ \cite{Gatti2025} & $0.273 \pm 0.029$ & $0.817 \pm 0.021$ & $< -0.72$ \\[2pt]
        DES Y3 map-level CNN$^*$ \cite{Jeffrey2025} & $0.283^{+0.020}_{-0.027}$ & $0.804^{+0.025}_{-0.017}$ & $< -0.80$ \\[2pt]
        DES Y3 persistent homology$^*$ \cite{Prat2025} & $0.304 \pm 0.037$ & $0.821 \pm 0.018$ & $< -0.75$ \\[2pt]
        DES Y3 hybrid statistics$^*$ \cite{Williamson2026} & $0.325 \pm 0.024$ & $0.808 \pm 0.017$ & $< -0.766$ \\[2pt]
        DES Y6 3$\times$2pt$^*$ \cite{DESY6_results} & $0.325^{+0.032}_{-0.035}$ & $0.781^{+0.021}_{-0.020}$ & $-1.12^{+0.26}_{-0.20}$ \\[2pt]
        \hline
        KiDS-1000 cosmic shear \cite{Troester2021} & $0.34^{+0.10}_{-0.10}$ & $0.743^{+0.049}_{-0.052}$ & $-1.02^{+0.45}_{-0.81}$ \\[2pt]
        KiDS-1000 3$\times$2pt \cite{Troester2021} & $0.302^{+0.017}_{-0.018}$ & $0.760^{+0.020}_{-0.021}$ & $-1.03^{+0.10}_{-0.13}$ \\[2pt]
        KiDS-Legacy cosmic shear$^*$ \cite{Reischke2026} & $0.332 \pm 0.059$ & $0.809 \pm 0.041$ & $-1.10^{+0.45}_{-0.47}$ \\[2pt]
        Planck 2018 + BAO \cite{Planck2018} & $0.304^{+0.011}_{-0.012}$ & $0.826^{+0.010}_{-0.011}$ & $-1.04 \pm 0.05$ \\[2pt]
        DESI DR2 BAO+CMB+SNe \cite{DESI_DR2} & $0.305 \pm 0.005$ & $0.814^{+0.008}_{-0.007}$ & $-0.995 \pm 0.023$ \\
        \hline
    \end{tabular}
    \caption{The \wcdm constraints on \Om, \Seight and \wzero shown in Figure~\ref{fig:literature_comparison}. We quote the median and 16th--84th percentiles from the public chains and, for the rows marked with $^*$, the published marginal mean and 68\% interval. For the analyses with a non-phantom prior, the published 68\% upper limit on \wzero is given.}
    \label{tab:literature_values}
\end{table}

\begin{table}[H]
    \centering
    \scriptsize
    \setlength{\tabcolsep}{3pt}
    \resizebox{\ifdim\width>\linewidth\linewidth\else\width\fi}{!}{%
    \begin{tabular}{|l|l|l|l|l|l|l|l|}
        \hline
        \textbf{Analysis} & \textbf{Data} & \textbf{Likelihood} & \textbf{IA} & \textbf{Baryons} & \textbf{$P(k)$ Ampl.} & $\sum m_\nu$ & \wzero \textbf{prior} \\
        \hline
        This work & \xipm, \zetapm & SBI, \texttt{CosmoGrid} & NLA (1) & $(\Mc, \nu)$ & flat \sigeight & fixed & $[-2, -1/3]^{\dagger}$ \\
        \hline
        cosmic shear & \xipm + SR & analytic & TATT (5) & scale cuts & flat $A_s$ & marg. & $[-2, -1/3]$ \\
        3$\times$2pt & 3$\times$2pt + SR & analytic & TATT (5) & scale cuts & flat $A_s$ & marg. & $[-2, -1/3]$ \\
        2+3pt shear$^*$ & \xipm, $\langle M_{ap}^3 \rangle$ & analytic & as DES Y3 & scale cuts & flat $A_s$ & marg. & $[-2, -1/3]$ \\
        moments+wavelets$^*$ & map statistics & SBI, GS & NLA (2) & scale cuts & flat \sigeight & marg. & $(-1, -1/3)$ \\
        map-level CNN$^*$ & $C_\ell$ + CNN & SBI, GS & NLA (2) & scale cuts & flat \sigeight & marg. & $(-1, -1/3)$ \\
        persistent homology$^*$ & Betti numbers & SBI, GS & NLA (2) & scale cuts & flat \sigeight & marg. & $(-1, -1/3)$ \\
        hybrid statistics$^*$ & $C_\ell$ + neural & SBI, GS & NLA (2) & scale cuts & flat \sigeight & marg. & $(-1, -1/3)$ \\
        Y6 3$\times$2pt$^*$ & 3$\times$2pt & analytic & TATT (5) & scale cuts & flat $A_s$ & marg. & $[-2, -1/3]$ \\
        \hline
        KiDS-1000 & band powers & analytic & NLA (1) & HMCode16 & flat \Seight & fixed & $[-3, -1/3]$ \\
        KiDS-1000 3$\times$2pt & + clustering & analytic & NLA (1) & HMCode16 & flat \Seight & fixed & $[-3, -1/3]$ \\
        KiDS-Legacy$^*$ & COSEBIs & analytic & mass-dep. & HMCode20 & flat \Seight & marg. & $[-3, 0]$ \\
        Planck 2018 + BAO & CMB, BAO & analytic & -- & -- & flat $\ln A_s$ & fixed & $[-3, 1]$ \\
        DESI DR2 & BAO, CMB, SNe & analytic & -- & -- & flat $\ln A_s$ & fixed & $[-3, 1]$ \\
        \hline
    \end{tabular}}
    \caption{Modelling choices of the analyses shown in Figure~\ref{fig:literature_comparison}. GS denotes the Gower Street simulations, SR the DES Y3 shear ratios, 3$\times$2pt the combination of \xipm, $\gamma_t$ and $w(\theta)$, `+ clustering' the BOSS and 2dFLenS clustering, SNe the Pantheon+ supernovae and `mass-dep.' the mass-dependent IA model of KiDS-Legacy; `as DES Y3' follows the IA setup of the DES Y3 \xipm analyses. The number of free IA parameters is given in parentheses, and $\sum m_\nu$ is either fixed at $0.06$\,eV or marginalised. Rows marked with $^*$ have no public chain; $^{\dagger}$ includes the \Om--\wzero restriction of Table~\ref{tab:prior_range}.}
    \label{tab:literature_setups}
\end{table}

The DES block contains analyses of the same Y3 source catalogue with different summary statistics and models. Any differences between these rows are therefore due to the methods and not due to the data. The DES Y6 3$\times$2pt result \cite{DESY6_results} is an update of the Y3 results and is not independent of them. It uses a new source catalogue and a new lens sample, and we quote its fiducial result. The DES Y3 \xipm + $\langle M_{ap}^3 \rangle$ analysis \cite{Gomes2025_des_a, Gomes2025_des_b} is the closest analogue of our data vector, as it combines second- and third-order shear statistics, but using an analytical model and an explicit likelihood of Ref.~\cite{Percival2022}. Its \wzero prior follows the DES Y3 \xipm analyses and allows $\wzero < -1$, which gives a two-sided constraint ($\wzero = -1.39 \pm 0.31$). The four map-based SBI analyses \cite{Gatti2025, Jeffrey2025, Prat2025, Williamson2026}, on the other hand, all use the Gower Street simulation suite, a $\nu$\wcdm parameter space and a non-phantom \wzero prior that covers half our range. Because of this, they only quote upper limits (from $<-0.72$ to $<-0.8$, compatible with our result).

Our improvement over \xipm (Table~\ref{tab:improvement}) is obtained on comparatively large angular scales, as the global and local \xipm start at $15'$ and the smallest aperture radius is $50'$. The map-based DES Y3 higher-order analyses instead include much smaller scales in their forward model, working on \texttt{HEALPIX} maps with a pixel scale of $6.9'$ up to $\ell \approx 1000$ \cite{Gatti2025, Jeffrey2025, Prat2025, Williamson2026} or, with a smoothing that depends on the tomographic bin, up to $\ell \approx 600$--$1400$ \cite{Thomsen2026}. For this, they rely on dedicated robustness tests against baryonic feedback or, in Ref.~\cite{Thomsen2026}, on the marginalisation over the baryonification parameters. In Paper~1 we instead chose a larger scale cut that excludes the scales most affected by baryonic feedback and resolution effects from the simulations, at the cost of some non-Gaussian information. Our improvement over \xipm is therefore not due to additional small-scale information, but comes from the third-order information on scales where the forward model mildly depends on baryonic feedback, which can also be seen in the weak constraint on \Mc (Section~\ref{sec:constraints_systematics}).

The independent block contains results from the KiDS-1000 beyond-\lcdm analysis \cite{Troester2021}, which uses band powers instead of COSEBIs, both for cosmic shear alone and for 3$\times$2pt with BOSS and 2dFLenS clustering, and KiDS-Legacy from the extended-cosmology analysis \cite{Reischke2026} of the fiducial COSEBIs setup \cite{Kids_results}. We show Planck 2018 \cite{Planck2018} together with BAO, because the CMB alone cannot constrain \wzero due to the geometric degeneracy. DESI DR2 BAO \cite{DESI_DR2} is shown in combination with the CMB (Planck PR4 CamSpec and CMB lensing) and the Pantheon+ supernovae sample that does not include DES data.

\subsection{Tension with independent data}\label{sec:comparison_tension}

We quantify the tension with the independent (external) data with the parameter-shift probability introduced in \cite{Raveri2019, Raveri2021}. It is based on the distribution of the parameter difference between two independent posteriors,
\begin{equation}\label{eq:param_shift_dist}
    \mathcal{P}(\Delta\theta) = \int \mathcal{P}_{\rm this\,work}(\theta)\, \mathcal{P}_{\rm ext}(\theta - \Delta\theta)\, \mathrm{d}\theta ,
\end{equation}
which is integrated above its zero-shift iso-contour,
\begin{equation}\label{eq:param_shift}
    \Delta = \int_{\mathcal{P}(\Delta\theta) > \mathcal{P}(0)} \mathcal{P}(\Delta\theta)\, \mathrm{d}\Delta\theta ,
\end{equation}
and converted to an effective significance $n_\sigma = \sqrt{2}\, \mathrm{Erf}^{-1}(\Delta)$. We compute it with the kernel density estimate of \texttt{tensiometer} \cite{Raveri2021, tensiometer} on our chains and each public chain, both in the (\Om, \Seight) plane and in the full (\Om, \Seight, \wzero) space, and list the resulting shifts for the three main analyses in Table~\ref{tab:tension}. For the joint analysis, the shifts are below $1\sigma$ for both KiDS-1000 results and reach $2.4\sigma$ for Planck 2018 + BAO and $1.6\sigma$ for DESI DR2 in this three-dimensional parameter space. These larger shifts come from the (\Om, \wzero) plane and not from the \wzero marginal, for which the simple marginal posterior estimate described below gives $\le 0.2\sigma$. The \zetapm-only posterior, which is broad and centred at higher \Om, is compatible with all of them, while the \xipm-only posterior shows a similar $\approx 2\sigma$ three-dimensional shift with respect to Planck 2018 + BAO as the joint analysis. For KiDS-Legacy, which has no public chain, we use the simple marginal posterior estimate $|\Delta\theta| / \sqrt{\sigma_1^2 + \sigma_2^2}$ \cite{Lemos2021}, with the marginal means and symmetrised 68\% intervals, as published for KiDS-Legacy, and find shifts below $1\sigma$ (Table~\ref{tab:tension}). For results with a public chain, we use the medians instead.

Overall, we do not find a significant tension with any of the independent results. The \Seight of the joint analysis lies below the Planck 2018 + BAO and DESI DR2 values, which tends in the direction of the low value of\Seight found in weak lensing surveys compared to CMB analyses. However, we note that this should only be interpreted bearing in mind the different priors and parameter spaces used in the analyses, in particular the \wzero ranges.

\subsection{Comparison with DES Y3 analyses}\label{sec:comparison_des}

For the other DES analyses, we do not compute a tension metric, as these analyses use the same source catalogue and are therefore correlated with our results. Using the simple estimate above, the \Seight of the joint analysis lies within $0.3\sigma$ of DES Y3 3$\times$2pt, $1.3\sigma$ above DES Y3 cosmic shear and $0.6$--$1.2\sigma$ below the four map-based SBI analyses, which find $\Seight = 0.80$--$0.82$.
\begin{table}[H]
    \centering
    \begin{tabular}{|l|cc|cc|cc|}
        \hline
        & \multicolumn{2}{c|}{\xipm + \zetapm} & \multicolumn{2}{c|}{\xipm} & \multicolumn{2}{c|}{\zetapm} \\
        \textbf{External data} & 2D & 3D & 2D & 3D & 2D & 3D \\
        \hline
        KiDS-1000 cosmic shear & $0.4\sigma$ & $0.3\sigma$ & $0.6\sigma$ & $0.4\sigma$ & $0.2\sigma$ & $0.1\sigma$ \\
        KiDS-1000 3$\times$2pt  & $0.8\sigma$ & $0.7\sigma$ & $1.0\sigma$ & $0.8\sigma$ & $0.4\sigma$ & $0.1\sigma$ \\
        Planck 2018 + BAO       & $1.2\sigma$ & $2.4\sigma$ & $0.9\sigma$ & $2.2\sigma$ & $1.3\sigma$ & $1.1\sigma$ \\
        DESI DR2                & $1.0\sigma$ & $1.6\sigma$ & $0.7\sigma$ & $1.6\sigma$ & $0.9\sigma$ & $0.6\sigma$ \\
        \hline
    \end{tabular}
    \caption{Parameter-shift significance $n_\sigma$ \cite{Raveri2019, Raveri2021} between the three main analyses and the independent results with a public chain, in the (\Om, \Seight) plane (2D) and in the (\Om, \Seight, \wzero) space (3D). The three-dimensional Planck value has the largest uncertainty of all estimates ($\pm 0.2\sigma$), which is limited by the effective number of samples in the public chain, and the estimates for the KiDS-1000 chains have uncertainties of $\pm 0.5$--$1\sigma$ for the same reason. For KiDS-Legacy, which has no public chain, the simple estimate of Section~\ref{sec:comparison_tension} gives $0.5\sigma$ in \Seight, $0.8\sigma$ in \Om and $0.2\sigma$ in \wzero for the joint analysis.}
    \label{tab:tension}
\end{table}

Part of this difference can be explained by their non-phantom \wzero prior: in our posterior, \Seight and \Om are positively correlated with \wzero (Figure~\ref{fig:corner}), and restricting it to $\wzero > -1$ gives posterior means of $\Seight = 0.803$ and $\Om = 0.290$. The other analysis combining second- and third-order shear statistics, \xipm + $\langle M_{ap}^3 \rangle$ \cite{Gomes2025_des_b}, which also allows $\wzero < -1$, lies within $0.6\sigma$ in \Om and $1.0\sigma$ in \Seight.

\begin{figure}[H]
    \centering
    \includegraphics[width=0.49\linewidth]{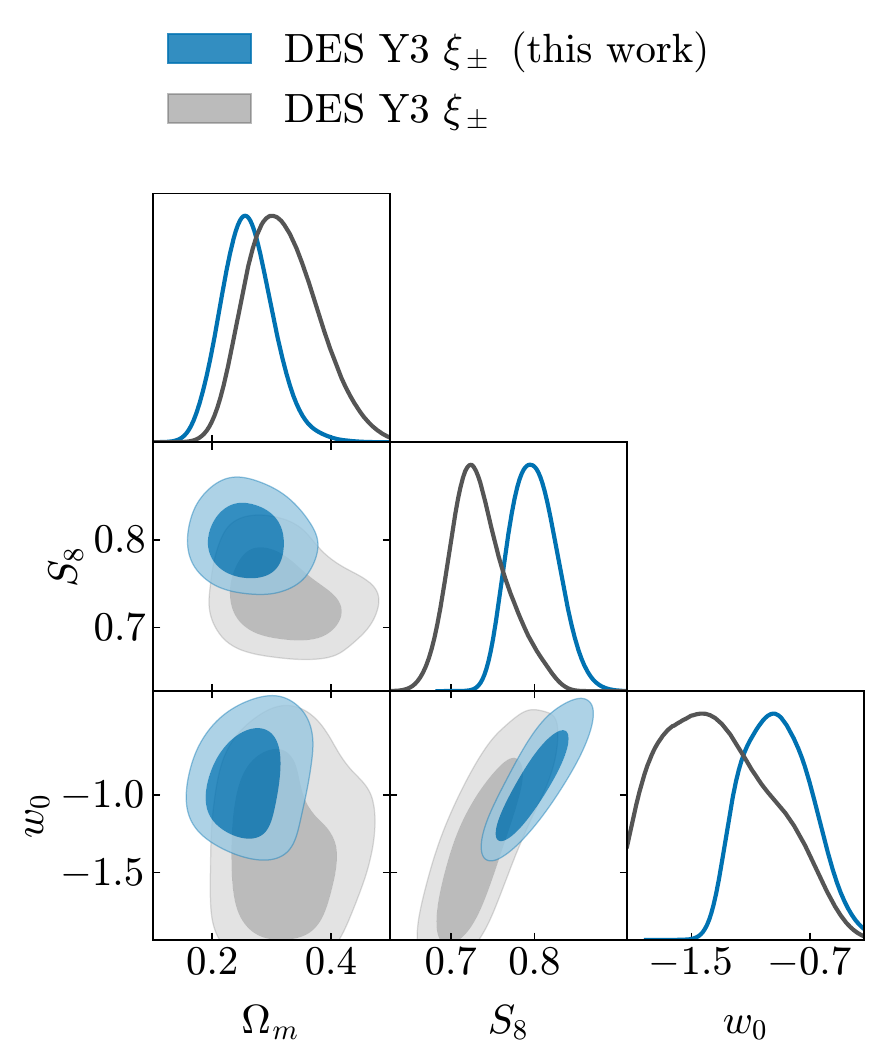}\hfill
    \includegraphics[width=0.49\linewidth]{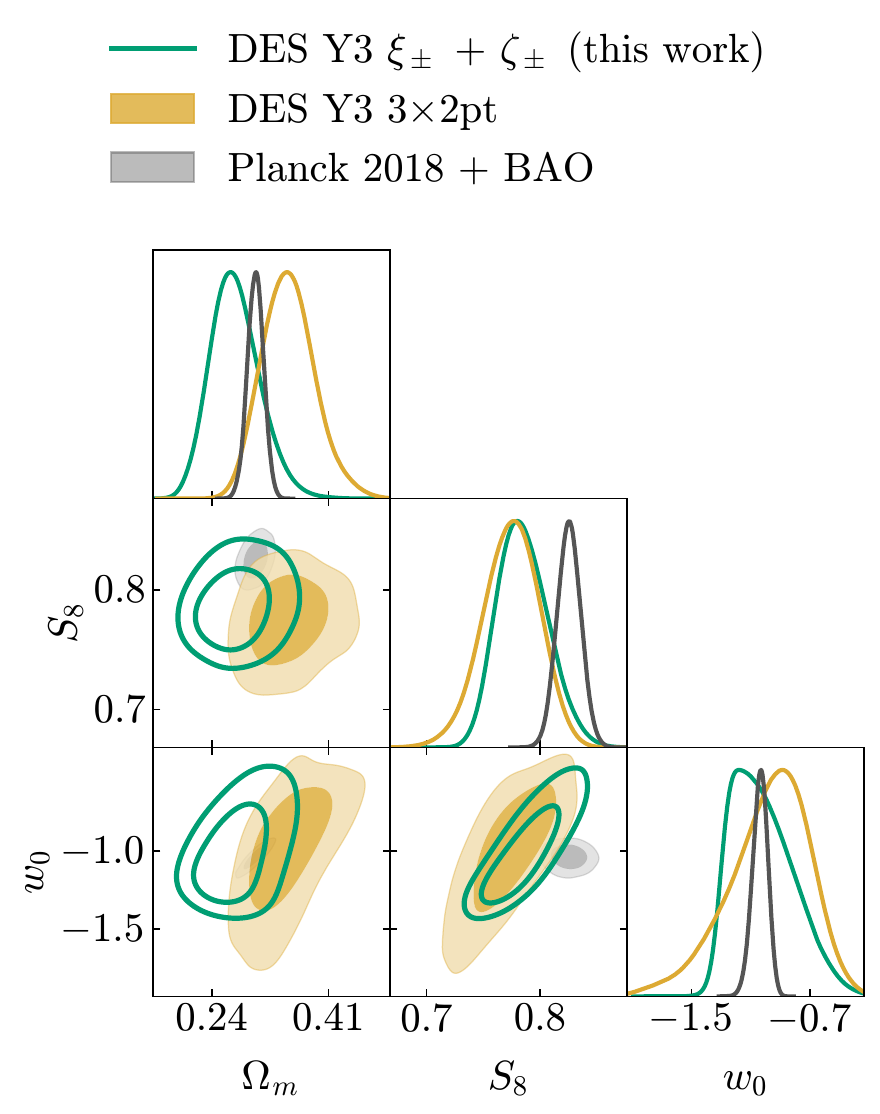}
    \begin{minipage}[t]{0.49\linewidth}
        \caption{Marginalised posterior distributions of \Om, \Seight and \wzero for the \xipm-only analysis of this work (blue), compared with the public \wcdm chain of the DES Y3 cosmic shear analysis \cite{Amon2022b, Secco2022a} (grey). Both analyses use the same source catalogue, so any difference is due to the methods and not the data. The 2D panels show the 68\% and 95\% credible contours.}
        \label{fig:corner_literature_2pcf}
    \end{minipage}\hfill
    \begin{minipage}[t]{0.49\linewidth}
        \caption{Same as Figure~\ref{fig:corner_literature_2pcf} but for the joint \xipm + \zetapm analysis (green, unfilled), compared with the public \wcdm chains of the DES Y3 3$\times$2pt analysis \cite{DESY3_results} (yellow) and of Planck 2018 + BAO \cite{Planck2018} (grey).}
        \label{fig:corner_literature_combined}
    \end{minipage}
\end{figure}

Figures~\ref{fig:corner_literature_2pcf} and \ref{fig:corner_literature_combined} show the \xipm-only posterior against the public DES Y3 cosmic shear \wcdm chain, and the joint analysis against the DES Y3 3$\times$2pt chain and Planck 2018 + BAO, in the \Om--\Seight--\wzero space. Figure~\ref{fig:corner_literature_2pcf} is a direct comparison of our \xipm-only analysis with the official DES Y3 cosmic shear result, which uses the same statistic and the same catalogue, so that the differences between the two are only in the modelling (Table~\ref{tab:literature_setups}) and in the \texttt{MOPED} compression. Our \xipm-only posterior is tighter, which is mainly due to the smaller scale cuts: the DES Y3 analysis removes the scales of \xipm on which baryonic feedback affects the analytical model, and for \xim these reach well above our lower limit of $15'$, while in our case the feedback is part of the forward model and marginalised over. The fixed neutrino mass and the one-parameter IA model also contribute to the tighter posterior. The IA amplitude of our joint analysis, $\Aia = 2.7^{+1.4}_{-1.9}$ (Section~\ref{sec:constraints_systematics}), is $1.2\sigma$ larger than the DES Y3 cosmic shear result $a_1 = 0.34^{+0.25}_{-0.23}$ \cite{Secco2022a}, obtained with the same NLA model without redshift evolution or shear ratios (their $a_1$ corresponds to our \Aia), and also larger than the values of the map-based SBI analyses ($0.37^{+0.30}_{-0.29}$ \cite{Gatti2025}, $0.21^{+0.17}_{-0.25}$ from the additional \Aia compression of Ref.~\cite{Jeffrey2025}, and $0.64 \pm 0.35$ \cite{Prat2025}), which use the NLA model with a free redshift evolution. Compared to all of these, our \Aia is, however, considerably less constrained, and the difference may be caused by the different priors, scale cuts and treatments of baryonic feedback. The \xipm-only posterior lies at the high-\Seight edge of the 68\% contour of DES Y3 cosmic shear, whose broad tail towards low \Seight causes the $1.3\sigma$ difference quoted above. In Figure~\ref{fig:corner_literature_combined}, the joint analysis overlaps with the 3$\times$2pt contour in \Seight and lies at lower \Om, and Planck 2018 + BAO lies in the high-\Om, high-\Seight corner of the joint 95\% contour, which is in the (\Om, \wzero) direction that is responsible for the $2.4\sigma$ shift. As for \Aia, the differences in \Om, \Seight and \wzero from the DES Y3 cosmic shear and 3$\times$2pt analyses may be due to modelling choices such as the different priors, scale cuts and treatments of baryonic feedback. In particular, \wzero is only weakly constrained and correlated with \Om and \Seight, so that differences in the \wzero prior volume, including the restriction in the \Om--\wzero plane (Section~\ref{sec:forward_model}), can shift the marginalised constraints through projection effects.

\FloatBarrier
\section{Conclusion}\label{sec:conclusions}

Maximising the scientific return from Stage-IV cosmic shear surveys like Euclid and LSST will require the analysis of higher-order statistics alongside \xipm. However, before such statistics can be applied at Stage-IV precision, it is important to demonstrate that they work on Stage-III survey data. This requires a realistic forward model that includes the relevant systematic effects, together with validation tests on blinded data.

In this work, we have presented the first cosmological analysis of the integrated 3-point correlation function on survey data. We have used \texttt{SBi3PCF}, the simulation-based inference framework developed and validated in Paper~1, for a blinded analysis of the DES Y3 weak lensing shape catalogue and obtained \wcdm posteriors for the cosmic shear statistics \xipm, \zetapm and their combination. Our main findings are as follows:

\begin{enumerate}[label=(\roman*)]
    \item We report the first detection of \zetapm on survey data, with a signal-to-noise ratio of $\mathrm{S/N} = 12.8$ for \zetapm alone and of 33.0 for the joint \xipm + \zetapm data vector (Section~\ref{sec:data}).
    \item All pre-unblinding tests, i.e.~coverage, accuracy of the learned likelihoods, internal consistency and goodness of fit of the compressed and the uncompressed data vector, were passed on the blinded measurement. The internal consistency tests between parts of the data vector and between the two statistics are calibrated on simulations instead of relying on raw $p$-values (Section~\ref{sec:validation}).
    \item For the joint \xipm + \zetapm analysis, we find $\Om = 0.272^{+0.038}_{-0.033}$, $\Seight = 0.784^{+0.024}_{-0.021}$ and $\wzero = -1.04^{+0.25}_{-0.19}$. We also find the intrinsic alignment amplitude $\Aia = 2.7^{+1.4}_{-1.9}$ and a weak constraint on baryonic feedback, $\log \Mc = 13.9^{+0.8}_{-1.1}$. Adding \zetapm improves the figure of merit in the \Om--\Seight--\wzero parameter space by a factor of $1.59$ over \xipm alone (Table~\ref{tab:improvement}), in line with the expectation from the simulated measurements of Paper~1. We note that it is obtained using only angular scales $\geq 15'$ and aperture radii $\geq 50'$, meaning that we do not rely on the small scales that the map-based higher-order analyses use (Section~\ref{sec:constraints}).
    \item Our constraints agree with the other DES Y3 analyses of the same source catalogue and with the DES Y6 update, the \xipm-only analysis is consistent with the official DES Y3 cosmic shear result in a direct comparison, and we find no significant tension with independent data, with parameter shifts of at most $2.4\sigma$ for Planck 2018 + BAO (Section~\ref{sec:comparison}).
\end{enumerate}

The present analysis has some limitations. The forward model is based on a single simulation suite, and we could not test the recovery of the input cosmology on simulations from an independent N-body code. The prior is inherited from \texttt{CosmoGridV1} and spans only \wcdm at a fixed total neutrino mass $\sum m_\nu$, so that there is no corresponding \lcdm analysis. Intrinsic alignments are only infused with the NLA model, with no redshift dependence ($\eta = 0$), but we find no impact of \Aia on cosmology. The \texttt{MOPED} compression vectors are fixed at one fiducial cosmology, which can lead to a loss of information, but not to a bias (Section~\ref{sec:sbi}). Since the fiducial values lie within the 68\% credible intervals of the joint analysis, we expect this loss to be small.
The $\nu$ and $m_3$ directions of the learned likelihood are not reliable and were not used for the consistency criterion in Section~\ref{sec:validation}, while the cosmological coefficients show no tension. Finally, source clustering and reduced shear were left out of the forward model based on the tests of Paper~1, which found their effect negligible.

We have thus shown that \zetapm can be analysed on Stage-III survey data with a fully simulation-based approach, an important step towards Stage-IV surveys like Euclid and Rubin LSST. At their precision, the map-level forward model will need tidal-field-based alignment models \cite{joachim_TATT_2025}, the inclusion of photometric redshift outliers \cite{Mill2025} and a treatment of baryonic feedback beyond two parameters \cite{Lucie-Smith_2025}, as well as larger simulation suites or emulators of the data vector to reduce the noise in the learned likelihood due to the finite number of simulations \cite{Homer2025}. These extensions can build on the pipeline presented here, as can joint analyses with other higher-order statistics of the shear maps or with statistics of photometric galaxy clustering maps. Together with the comparison with the analytic-likelihood \zetapm analysis of the same data (Halder et al.~in preparation), they can contribute to the growing efforts in the weak lensing community towards robust HOS analyses that more fully exploit the non-Gaussian information encoded in the large-scale structure of the Universe.

\FloatBarrier
\acknowledgments
We would like to thank Lina Castiblanco, Oliver Friedrich, Marco Gatti, Daniel Gruen, Joachim Harnois-Déraps, Jed Homer, Mike Jarvis, Tomasz Kacprzak, Kai Lehman, Nils Schöneberg, Arne Thomsen, Simone Vinciguerra, and especially the late Stella Seitz for fruitful discussions at various stages of this project. We are grateful to Tomasz Kacprzak and the \texttt{CosmoGridV1} team for making the simulations publicly available. We acknowledge support from the Excellence Cluster ORIGINS, which is funded by the Deutsche Forschungsgemeinschaft (DFG, German Research Foundation) under Germany’s Excellence Strategy -- EXC-2094-390783311. Some of the numerical calculations have been carried out on the ORIGINS computing facilities of the Computational Center for Particle and Astrophysics (C2PAP). DG and CU were supported by the European Union (ERC StG, LSS\_BeyondAverage, 101075919). DG acknowledges support from the German Academic Exchange Service (DAAD) through a Research Scholarship for Doctoral Candidates, and thanks the Institute of Astronomy, University of Cambridge, for hosting him as a visiting student. AH has been supported by funding from the European Research Council (ERC) under the European Union’s Horizon 2020 research and innovation programmes (grant agreement no. 101018897 CosmicExplorer). ZG acknowledges support via the funding from the U.S. Department of Energy Cosmic Frontier program, grant DE-SC002599.

This project used public archival data from the Dark Energy Survey (DES). Funding for the DES Projects has been provided by the U.S. Department of Energy, the U.S. National Science Foundation, the Ministry of Science and Education of Spain, the Science and Technology Facilities Council of the United Kingdom, the Higher Education Funding Council for England, the National Center for Supercomputing Applications at the University of Illinois at Urbana--Champaign, the Kavli Institute of Cosmological Physics at the University of Chicago, the Center for Cosmology and Astro-Particle Physics at the Ohio State University, the Mitchell Institute for Fundamental Physics and Astronomy at Texas A\&M University, Financiadora de Estudos e Projetos, Fundação Carlos Chagas Filho de Amparo à Pesquisa do Estado do Rio de Janeiro, Conselho Nacional de Desenvolvimento Científico e Tecnológico and the Ministério da Ciência, Tecnologia e Inovação, the Deutsche Forschungsgemeinschaft and the Collaborating Institutions in the Dark Energy Survey. The Collaborating Institutions are Argonne National Laboratory, the University of California at Santa Cruz, the University of Cambridge, Centro de Investigaciones Enérgeticas, Medioambientales y Tecnológicas--Madrid, the University of Chicago, University College London, the DES-Brazil Consortium, the University of Edinburgh, the Eidgenössische Technische Hochschule (ETH) Zürich, Fermi National Accelerator Laboratory, the University of Illinois at Urbana-Champaign, the Institut de Ciències de l'Espai (IEEC/CSIC), the Institut de Física d'Altes Energies, Lawrence Berkeley National Laboratory, the Ludwig-Maximilians Universität München and the associated Excellence Cluster Universe, the University of Michigan, the National Optical Astronomy Observatory, the University of Nottingham, The Ohio State University, the OzDES Membership Consortium, the University of Pennsylvania, the University of Portsmouth, SLAC National Accelerator Laboratory, Stanford University, the University of Sussex, and Texas A\&M University. Based in part on observations at Cerro Tololo Inter-American Observatory, National Optical Astronomy Observatory, which is operated by the Association of Universities for Research in Astronomy (AURA) under a cooperative agreement with the National Science Foundation.

During the preparation of this work, the authors used Claude Code and several Claude models (Anthropic, 2026 versions) to assist with writing analysis and plotting code and to improve the language and readability of parts of the text. All AI-assisted code was reviewed and validated by the authors, and all figures are produced directly from the measured data and simulations. The authors reviewed and edited all AI-assisted text and take full responsibility for the content of the manuscript.

This work made use of \texttt{numpy} \cite{Harris2020}, \texttt{scipy} \cite{Virtanen2020}, \texttt{matplotlib} \cite{Hunter2007}, \texttt{ChainConsumer} \cite{Hinton2016}, \texttt{PyTorch} \cite{Paszke2019}, \texttt{sbi} \cite{sbi2020}, \texttt{zuko} \cite{Rozet2024}, \texttt{optuna} \cite{optuna_2019}, \texttt{TARP} \cite{Lemos2023} and \texttt{TreeCorr} \cite{Jarvis2004, Jarvis2019}.

\medskip\noindent\textbf{Data availability:} The full analysis code is publicly available at \href{https://github.com/D-Gebauer/SBi3PCFII}{\texttt{SBi3PCFII} \faGithub}, and the data products required to reproduce every figure and quoted number of this paper are archived on Zenodo (\url{https://doi.org/10.5281/zenodo.22018495}). The training and covariance sets can be reconstructed from the public \texttt{CosmoGridV1} simulations \cite{Kacprzak2023}, with the weak lensing maps built using \texttt{UFalcon2} \cite{Sgier2019, Sgier2021, Reeves2024}, the \xipm measurements performed with \texttt{TreeCorr} \cite{Jarvis2004, Jarvis2019} and the \zetapm measurements with \href{https://github.com/D-Gebauer/PixPoint}{\texttt{PixPoint} \faGithub}, following the procedure of Paper~1. These sets can also be shared by the authors upon request.

\appendix
\FloatBarrier
\section{Pre-unblinding tests}\label{sec:app_validation}

In this appendix, we describe the pre-unblinding tests that are summarised in Section~\ref{sec:validation}. The first two tests are carried out on the simulated measurements of the test set, the remaining three on the blinded data vector, and all of them had to be passed before unblinding. Where the DES Y3 measurement enters, the figures show the unblinded measurement and the corresponding blinded results are quoted in the text.

\subsection{Coverage tests}\label{sec:coverage_tests}

We verify that the posteriors of all eight inference pipelines used in this work are statistically well calibrated with the \texttt{TARP} \cite{Lemos2023} coverage test, as in Paper~1. The test is evaluated on $N = 1250$ test simulations that are distributed following the prior and were excluded from training (Paper~1 used 500), with the reference points of \texttt{TARP} drawn uniformly over the prior box. The empirical coverage probability ECP$(\alpha)$ is the fraction of the test inferences for which the true parameter values are enclosed within the $\alpha$-level credible interval, and the binomial band $\pm \sqrt{\alpha(1-\alpha)/N}$ around the diagonal gives its expected scatter for $N$ simulations.

\begin{figure}[H]
    \centering
    \includegraphics[width=0.8\linewidth]{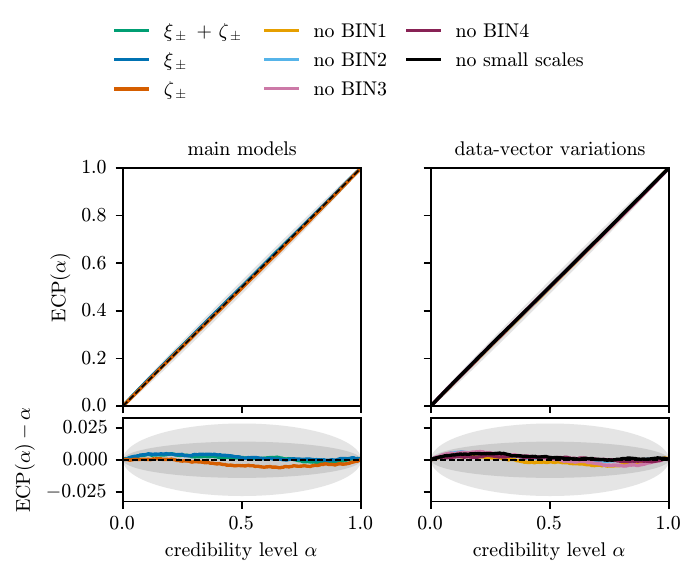}
    \caption{Coverage test with \texttt{TARP} \cite{Lemos2023} on 1250 test simulations for the three main models (left) and the five data-vector variations (right). The upper panels show ECP$(\alpha)$ as a function of the credibility level $\alpha$, and the lower panels show the residual ECP$(\alpha) - \alpha$. The black dashed line indicates perfect coverage, and the grey bands show the $1\,(2)\,\sigma$ binomial uncertainty.}
    \label{fig:tarp_coverage}
\end{figure}

Figure~\ref{fig:tarp_coverage} shows that all curves lie within the $1\,(2)\,\sigma$ binomial band around the one-to-one line, so that the posteriors of the main models and of the data-vector variations are statistically well calibrated. The largest deviation is $|{\rm ECP}(\alpha) - \alpha| = 0.007$ for `no BIN3', compared to a $2\sigma$ binomial band of $0.028$ at $\alpha = 0.5$, and the mean deviations lie between $-0.003$ and $+0.002$, i.e.~the curves scatter around the diagonal without a systematic offset.

We note that passing such a coverage test is a necessary condition for accurate posteriors, but not a sufficient one. The equivalence between exact expected coverage and posterior accuracy shown in Ref.~\cite{Lemos2023} only holds in the limit of infinitely many test simulations and every possible reference point, so that a passed test on a finite test set limits, but does not exclude, a residual miscalibration of the posteriors.

\subsection{Accuracy of the learned likelihoods}\label{sec:app_likelihoods}

\begin{figure}[H]
    \centering
    \includegraphics[width=\linewidth]{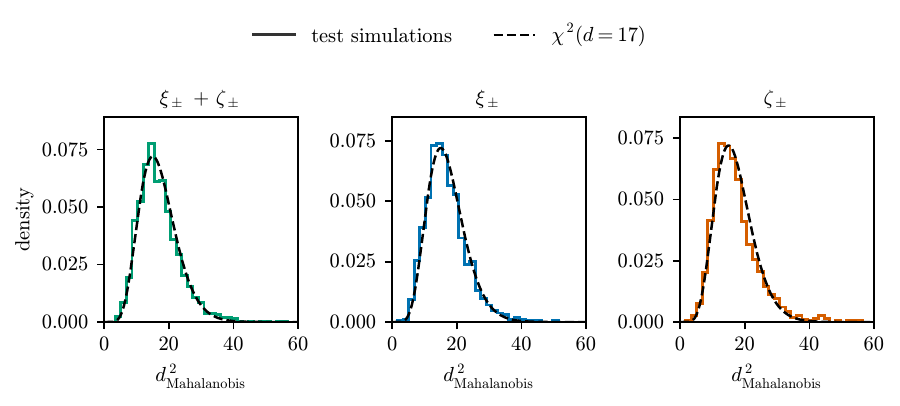}
    \caption{Histograms of the $\chi^2$ (squared Mahalanobis distance) of the 1250 test data vectors with respect to their learned likelihood $p(x \,|\, \theta_{\rm true})$ for the joint (left), \xipm (middle) and \zetapm (right) pipelines. The black dashed line shows the $\chi^2(d = 17)$ distribution expected for an accurately learned likelihood.}
    \label{fig:likelihood_mahalanobis}
\end{figure}

While the coverage test checks the posteriors, we also test the learned likelihoods directly. For this, we compute the squared Mahalanobis distance of each test data vector with respect to its own learned likelihood $p(x \,|\, \theta_{\rm true})$, which for a Gaussian likelihood is exactly the $\chi^2$, and compare its distribution over the test set with the $\chi^2$ distribution expected for the $d = 17$ elements of the compressed data vector. This checks whether the widths of the learned likelihoods are correct. The distance is evaluated in the space of the standardised \texttt{MOPED} coefficients, with the mean and covariance of each likelihood estimated from 1000 draws of the flow and the inverse covariance corrected with the Hartlap factor \cite{Hartlap2007}. Figure~\ref{fig:likelihood_mahalanobis} shows the resulting distributions for the three main analyses. The medians per degree of freedom are $0.93$--$0.94$, which shows that the learned likelihoods are not systematically too wide or too narrow, while the means ($1.18$--$1.28$) are raised by a tail of poorly fit cosmologies: $7$--$9\%$ of the test cases lie above the $95\%$ quantile of $\chi^2(17)$, and the 9--12 cases with $\chi^2 > 100$ (outside of the plotted range) all lie at the corners of the prior, where the training set is sparsest.

We compute the same statistic for the DES Y3 measurement with respect to the learned likelihood at its MAP and find $17.1$ (joint), $17.8$ (\xipm) and $20.0$ (\zetapm) for 17 degrees of freedom, corresponding to the 56th, 60th and 73rd percentile of $\chi^2(17)$. For the blinded data vector, the values were $20.5$, $21.1$ and $24.7$. The learned likelihoods thus describe the DES Y3 measurement as well as a typical mock.

\subsection{Internal consistency tests}\label{sec:app_consistency}

\begin{figure}[H]
    \centering
    \includegraphics[width=\linewidth]{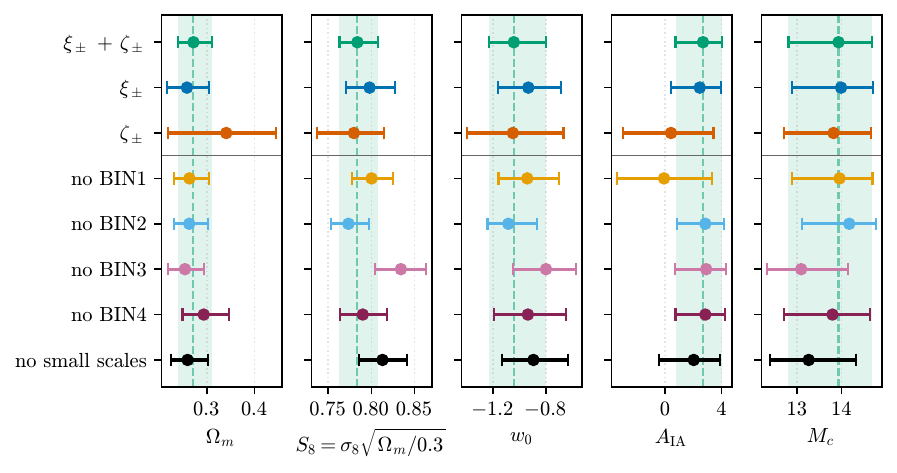}
    \caption{Posterior median and 68\% credible interval of the DES Y3 measurement for all analysis variations. The shaded band marks the joint \xipm + \zetapm interval as reference.}
    \label{fig:observation_model_variations}
\end{figure}

Figure~\ref{fig:observation_model_variations} shows the DES Y3 posteriors for the three main analyses and the five  variations of the joint \xipm + \zetapm data vector, in which either one tomographic bin or the two smallest angular bins are removed. All variations are consistent with the joint posterior. To quantify this, Table~\ref{tab:ablation_shifts} lists the shifts of the posterior medians of the five variations relative to the joint analysis in units of the 68\% half-width of the joint posterior,

\begin{equation}
    \epsilon = \frac{\theta^{\text{var}}_{\text{median}} - \theta^{\text{joint}}_{\text{median}}}{\frac{1}{2} \left| \theta^{\text{joint}}_{84\%} - \theta^{\text{joint}}_{16\%}\right|},
\end{equation}
which is the metric of Section~5.3 of Paper~1 with the joint median in place of the true value. Positive values lie above the joint median. The largest shift is the upward \Seight shift of `no BIN3' ($+2.2$), followed by `no small scales' ($+1.3$); all other shifts in the cosmological parameters are at most $1.1$, and the largest shift in a systematic parameter is $-1.7$ in \Aia for `no BIN1'. We note that these should not be read as significances, since removing part of the data changes both the position and the width of the posterior. The quantitative consistency test is the posterior predictive test of Appendix~\ref{sec:app_ppd}.

Table~\ref{tab:ablation_constraints} lists the marginalised constraints shown in Figure~\ref{fig:observation_model_variations}. Relative to the joint analysis, the variations widen the constraints by at most a factor of $1.35$ in \Om and $1.26$ in \wzero (`no BIN4') and $1.33$ in \Seight (`no BIN3'), and `no BIN1' loses the constraint on \Aia ($\times 2.0$, with a median compatible with zero), as the lowest-redshift bin carries most of the intrinsic alignment signal. On the blinded measurement, we found the same pattern of shifts (`no BIN3' $+2.29$ and `no small scales' $+1.28$ in \Seight, \Aia $-1.66$ for `no BIN1'), so that the outcome of this test is unchanged by unblinding.

The shifts of the variations are also consistent with the other tests in this appendix. The upward shift in \Seight for `no BIN3' matches the $-2.1\sigma$ deviation of the $m_3$ coefficient in Figure~\ref{fig:obs_moped_coefficients} and the lowest $\tilde{p}(T)$ of the posterior predictive tests below (Table~\ref{tab:ppd_consistency}), all indicating that the amplitude of the third tomographic bin is slightly lower than what the joint model predicts. Similarly, removing the two smallest angular bins raises \Seight ($+1.3$) and lowers \Mc ($-0.7$), i.e.~the scales between $15'$ and $30'$ prefer a slightly lower amplitude, as one would expect from baryonic suppression. As discussed in Section~\ref{sec:validation}, both shifts move along the \Seight--\wzero degeneracy seen in Figure~\ref{fig:corner}, and we consider them to be a single shift rather than two separate ones.

\subsection{Likelihood of the observed data vector}\label{sec:app_loglike}

\begin{figure}[H]
    \centering
    \includegraphics[width=\linewidth]{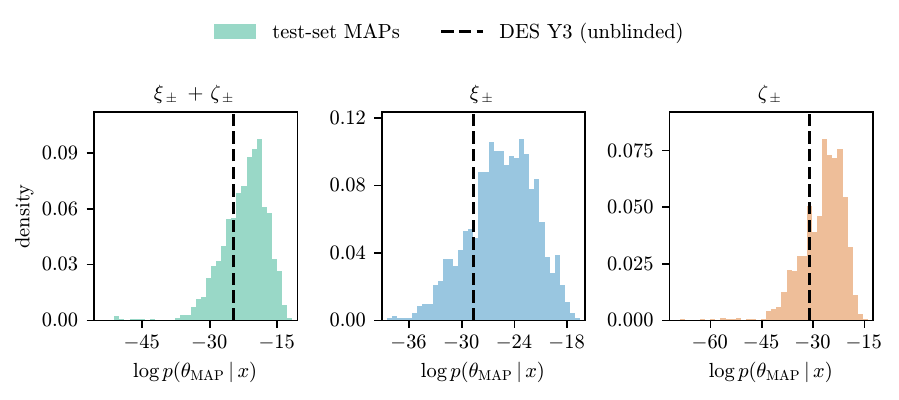}
    \caption{Log-posterior at the MAP of the DES Y3 data vector (black dashed) compared to its distribution for the 1250 test simulations, for the joint (left), \xipm (middle) and \zetapm (right) pipelines. A few test cases lie below the plotted range.}
    \label{fig:map_loglike_check}
\end{figure}

Figure~\ref{fig:map_loglike_check} compares the log-posterior at the MAP of the DES Y3 measurement with the distribution of the same quantity over the test set, for which the MAP of each of the 1250 test simulations is evaluated from its own posterior in the same way as for the measurement (Section~\ref{sec:sbi}). Since the prior is flat within its support, the differences within a panel are differences in the log-likelihood. The measurement lies at the 28th (joint), 20th (\xipm) and 24th (\zetapm) percentile of the test-set distributions, and the blinded data vector at the slightly lower 23rd, 16th and 21st percentile. The observed data vector is therefore a typical realisation of the forward model, and we find no indication that the forward model does not describe the data.
\begin{table}[H]
    \centering
    \begin{tabular}{|l|cccc|}
        \hline
        \textbf{Analysis} & \Om & \Seight & \wzero & \Aia \\
        \hline
        \xipm + \zetapm    & $0.272^{+0.038}_{-0.033}$ & $0.784^{+0.024}_{-0.021}$ & $-1.04^{+0.25}_{-0.19}$ & $2.7^{+1.4}_{-1.9}$ \\[2pt]
        no BIN1         & $0.264^{+0.041}_{-0.033}$ & $0.800^{+0.025}_{-0.023}$ & $-0.94^{+0.24}_{-0.22}$ & $-0.1^{+3.4}_{-3.3}$ \\[2pt]
        no BIN2         & $0.263^{+0.040}_{-0.032}$ & $0.773^{+0.024}_{-0.020}$ & $-1.09^{+0.22}_{-0.16}$ & $2.9^{+1.3}_{-2.0}$ \\[2pt]
        no BIN3         & $0.254^{+0.041}_{-0.036}$ & $0.834^{+0.029}_{-0.030}$ & $-0.80^{+0.23}_{-0.25}$ & $2.9^{+1.4}_{-2.2}$ \\[2pt]
        no BIN4         & $0.294^{+0.052}_{-0.044}$ & $0.790^{+0.028}_{-0.026}$ & $-0.94^{+0.29}_{-0.26}$ & $2.9^{+1.4}_{-2.1}$ \\[2pt]
        no small scales & $0.260^{+0.044}_{-0.034}$ & $0.813^{+0.029}_{-0.027}$ & $-0.90^{+0.26}_{-0.24}$ & $2.0^{+1.8}_{-2.5}$ \\
        \hline
    \end{tabular}
    \caption{Marginalised constraints (median and 16th--84th percentiles) on \Om, \Seight, \wzero and \Aia from the unblinded DES Y3 measurement for the joint analysis and the five data-vector variations of Figure~\ref{fig:observation_model_variations}.}
    \label{tab:ablation_constraints}
\end{table}

\begin{table}[H]
    \centering
    \begin{tabular}{|l|ccccc|}
        \hline
        \textbf{Variation} & \Om & \Seight & \wzero & \Aia & \Mc \\
        \hline
        no BIN1         & $-0.23$ & $+0.73$ & $+0.47$ & $-1.67$ & $+0.02$ \\
        no BIN2         & $-0.25$ & $-0.47$ & $-0.19$ & $+0.10$ & $+0.25$ \\
        no BIN3         & $-0.50$ & $+2.25$ & $+1.13$ & $+0.14$ & $-0.89$ \\
        no BIN4         & $+0.60$ & $+0.28$ & $+0.49$ & $+0.10$ & $-0.15$ \\
        no small scales & $-0.34$ & $+1.30$ & $+0.69$ & $-0.40$ & $-0.71$ \\
        \hline
    \end{tabular}
    \caption{Shifts of the posterior medians of the five data-vector variations relative to the joint analysis for the unblinded DES Y3 measurement, in units of the 68\% half-width of the joint posterior. Positive values lie above the joint median.}
    \label{tab:ablation_shifts}
\end{table}

Figure~\ref{fig:obs_moped_coefficients} repeats this check for each of the nine \texttt{MOPED} coefficients corresponding to the model parameters, with the learned likelihood at the MAP represented by 1000 draws of the joint flow. We can see that the coefficients for \Om, \sigeight and \wzero deviate from the mean of the learned likelihood by at most $0.13\sigma$ and are therefore typical draws from it. All other coefficients, including the eight shear calibration and photo-$z$ coefficients not shown in the figure, lie within $1.3\sigma$, with the exception of the redshift evolution $\nu$ of the baryonification ($+2.9\sigma$) and the shear calibration bias $m_3$ ($-2.1\sigma$), to which we come back in the posterior predictive tests below.

\subsection{Posterior predictive consistency tests}\label{sec:app_ppd}
\begin{figure}[H]
    \centering
    \includegraphics[width=0.8\linewidth]{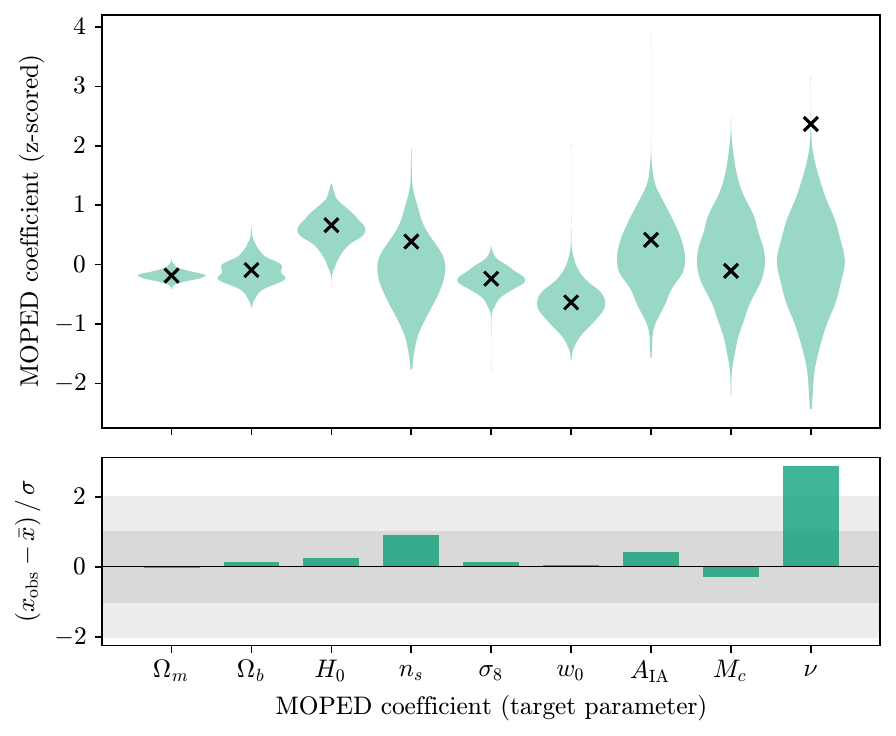}
    \caption{Top: The nine z-scored \texttt{MOPED} coefficients of the DES Y3 data vector (black crosses), compared to the learned likelihood $p(x \,|\, \theta_{\rm MAP})$ of the joint analysis (violins). Bottom: The deviation of the measured coefficients from the likelihood mean in units of its standard deviation. The shaded bands mark $\pm 1\sigma$ and $\pm 2\sigma$.}
    \label{fig:obs_moped_coefficients}
\end{figure}

Complementary to the comparison of the posteriors in Figure~\ref{fig:observation_model_variations}, we test the internal consistency of the data-vector variations, and of the two statistics with each other, with posterior predictive distribution (PPD) tests following the DES methodology of Ref.~\cite{Doux2021}. The idea behind these tests is to run the inference with part of the data removed, to predict the removed part from the resulting posterior, and to then compare this prediction to the actual measurement using a test statistic. The raw PPD $p$-values obtained in this way are not uniformly distributed, but skewed towards low values if the two parts of the data vector constrain different directions in parameter space. Following Ref.~\cite{Doux2021}, we therefore calibrate them on simulated realisations, which are consistent by construction: the calibrated $\tilde{p}$ is the fraction of the 800 fiducial noise realisations that have a larger test statistic than the measurement, and as in DES, we require $\tilde{p} > 0.01$ for consistency.

Ref.~\cite{Doux2021} uses an analytical Gaussian likelihood, draws realisations of the removed data from the PPD and compares them to the measurement with a $\chi^2$-like distance. Here, the likelihood is only available as the trained flow, so we use the likelihood of the actual measurement as the test statistic instead,
\begin{equation}\label{eq:ppd_T}
    T = -2\,\big\langle \log p_{\rm full}(x_{\rm full}\,|\,\tilde{\theta}) \big\rangle_{\tilde{\theta} = \theta_\text{post}} ,
\end{equation}
which is the joint flow evaluated at the full \texttt{MOPED} data vector and averaged over 2000 thinned chain samples $\tilde{\theta}$ from the posterior $\theta_\text{post}$ of the analysis with the reduced data vector. With this choice, we need neither realisations drawn from the PPD nor a distance measure in data space. As the normalisation of the flow is arbitrary, the absolute values of $T$ have no meaning on their own, but it cancels out in the calibration, and we only ever compare the calibrated $\tilde{p}$. We also note that the full data vector still contains the part of the data on which the posterior was conditioned, which biases the test toward agreement. This bias is present in the simulated measurements in the same way and is therefore also removed by the calibration. We compute $T$ for the five data-vector variations, for the two single-statistic analyses (the joint flow with the \xipm-only or the \zetapm-only posterior, with the other statistic as the removed data), and for the joint analysis itself, which amounts to a goodness-of-fit test.

For \xipm and \zetapm, we additionally perform two cross-prediction tests in which the data used for the posterior and the data at which the likelihood is evaluated do not overlap, which is the direct analogue of the consistency tests in Ref.~\cite{Doux2021}. Here, the flow of statistic $A$ is evaluated at its own data vector using parameters that were inferred from the other statistic $B$ alone,
\begin{equation}\label{eq:ppd_T_cross}
    T = -2\,\big\langle \log p_{A}(x_{A}\,|\,\tilde{\theta}) \big\rangle_{\tilde{\theta} = \theta_{\rm post,B}} ,
\end{equation}
with $\tilde{\theta}$ drawn from the posterior of statistic $B$. In this case, there is no bias towards agreement, since the data at which the likelihood is evaluated never enters the posterior, and the correlated noise of the two statistics, which are measured on the same maps, is again absorbed by the calibration. These tests differ from the single-statistic tests above only in how the removed statistic enters: there as one part of the jointly compressed \texttt{MOPED} coefficients, here through a separate compression.

\begin{figure}[H]
    \centering
    \includegraphics[width=\linewidth]{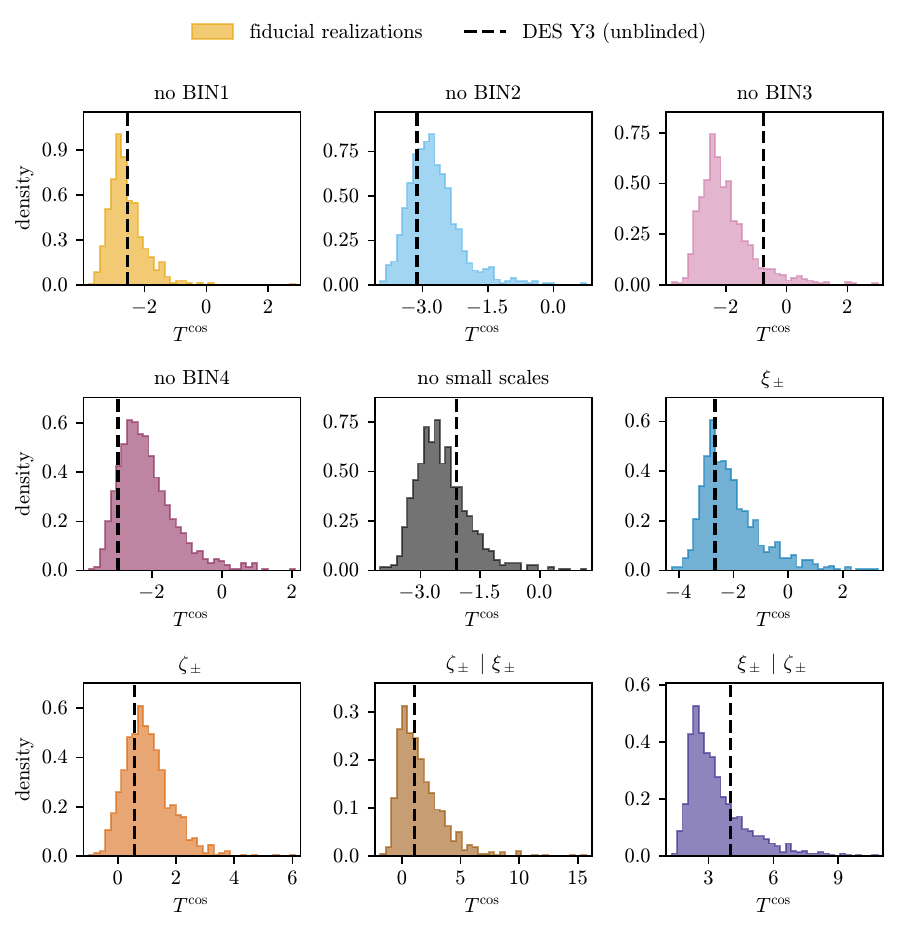}
    \caption{Calibration of the consistency statistic $T^{\rm cos}$ for the tests of Table~\ref{tab:ppd_consistency}. In each panel, the histogram shows the distribution of $T^{\rm cos}$ for the 800 fiducial noise realisations, and the dashed line marks the DES Y3 measurement.}
    \label{fig:ppd_consistency}
\end{figure}

Since the marginal distributions of the flow over a subset of the coefficients are not available, we also evaluate a Gaussian version of the statistic for the three coefficients corresponding to \Om, \sigeight and \wzero, with the mean and covariance estimated from one draw of the flow per chain sample. We denote this statistic by $T^{\rm cos}$ and show its calibration in Figure~\ref{fig:ppd_consistency}. It directly tests whether all parts of the data vector prefer the same values for the parameters that we report.

All tests on the cosmological coefficients are passed (Table~\ref{tab:ppd_consistency}) with $\tilde{p}(T^{\rm cos}) = 0.08$--$0.89$. The value is lowest for `no BIN3', where it is still eight times the threshold. The tests conditioning the full data vector on \xipm and \zetapm give $0.65$ and $0.70$ respectively, and the separate cross-predictions give $0.51$ and $0.24$. We can therefore conclude that the parameters inferred from any of the subparts of the data vector correctly predict the removed part of the measurement in \Om, \sigeight and \wzero.

For the statistic $T$ on the full data vector, the results are more marginal, with the lowest $\tilde{p}(T) = 0.001$ for `no BIN3' again, and even the goodness-of-fit test of the joint analysis gives only $0.021$, which is lower than the 28th percentile of the MAP log-posterior in Figure~\ref{fig:map_loglike_check} and the 56th percentile of the $\chi^2$ of Appendix~\ref{sec:app_likelihoods}. There are two reasons for that. Firstly, the fiducial realisations used for the calibration do not contain the poorly fit cosmologies at the corners of the prior, which widen the test-set distributions. And secondly, $T$ is evaluated with the full density fitted by the flow and not with its Gaussian approximation, in which a deviation in a single direction is diluted over the 17 degrees of freedom. Looking at individual coefficients, we find that the tension is concentrated in the $+2.9\sigma$ deviation in $\nu$ and the $-2.1\sigma$ deviation in $m_3$ of Figure~\ref{fig:obs_moped_coefficients}, while the cosmological coefficients show no tension. However, we do not base the consistency test on these two directions, as the learned likelihood is not reliable in them: during training we never pass the parameter value of $m_3$ to the flow, meaning it cannot learn the dependence of its corresponding \texttt{MOPED} coefficient, and since the data vector is barely sensitive to $\nu$, the likelihood in this direction is dominated by the prior. In these directions, the calibrated test does therefore not probe the internal consistency, but only how close the (noise-dominated) scatter of the observed data vector lies to the centre of the prior, and we base the consistency criterion on $T^{\rm cos}$.
\begin{table}[H]
    \centering
    \begin{tabular}{|l|cc|}
        \hline
        \textbf{Conditioning test} & $\tilde{p}(T^{\rm cos})$ & $\tilde{p}(T)$ \\
        \hline
        no BIN1 & 0.391 & 0.018 \\
        no BIN2 & 0.741 & 0.024 \\
        no BIN3 & 0.084 & 0.001 \\
        no BIN4 & 0.889 & 0.043 \\
        no small scales & 0.253 & 0.015 \\
        \hline
        \xipm & 0.651 & 0.064 \\
        \zetapm & 0.695 & 0.394 \\
        \hline
        \zetapm $|$ \xipm & 0.511 & 0.068 \\
        \xipm $|$ \zetapm & 0.235 & 0.438 \\
        \hline
        \xipm + \zetapm & 0.964 & 0.021 \\
        \hline
    \end{tabular}
    \caption{Calibrated $\tilde{p}$ values of the posterior predictive consistency tests for the DES Y3 measurement, for the statistic on the coefficients of \Om, \sigeight and \wzero ($T^{\rm cos}$) and on the full data vector ($T$). The cross-prediction tests are denoted `A $|$ B'.}
    \label{tab:ppd_consistency}
\end{table}

Comparing the two statistics further supports this interpretation. All tests that are conditioned on data containing \xipm have a marginal $\tilde{p}(T)$ ($0.001$--$0.07$), whereas the two tests conditioned on \zetapm have the highest values of all tests ($0.39$ and $0.44$), as one would expect if the $\nu$/$m_3$ tension enters through the \xipm part of the data vector. We note though, that the \zetapm-only posterior is much wider, and that conditioning on it is therefore a weaker test. The more meaningful direction is predicting \zetapm from \xipm, for which we find $\tilde{p}(T^{\rm cos}) = 0.51$ and $\tilde{p}(T) = 0.068$. On the blinded data vector, the results for $T^{\rm cos}$ were the same, with all tests passing in both cases, while the values of $\tilde{p}(T)$ were lower throughout (e.g.~`no BIN1' $0.005$ and `no small scales' $0.006$), meaning the $\nu$/$m_3$ tension is slightly weaker in the unblinded data vector, which is consistent with the MAP percentiles above.

\FloatBarrier
\section{Full Posterior}\label{sec:full_posterior}

\begin{figure}[H]
    \centering
    \includegraphics[width=\linewidth]{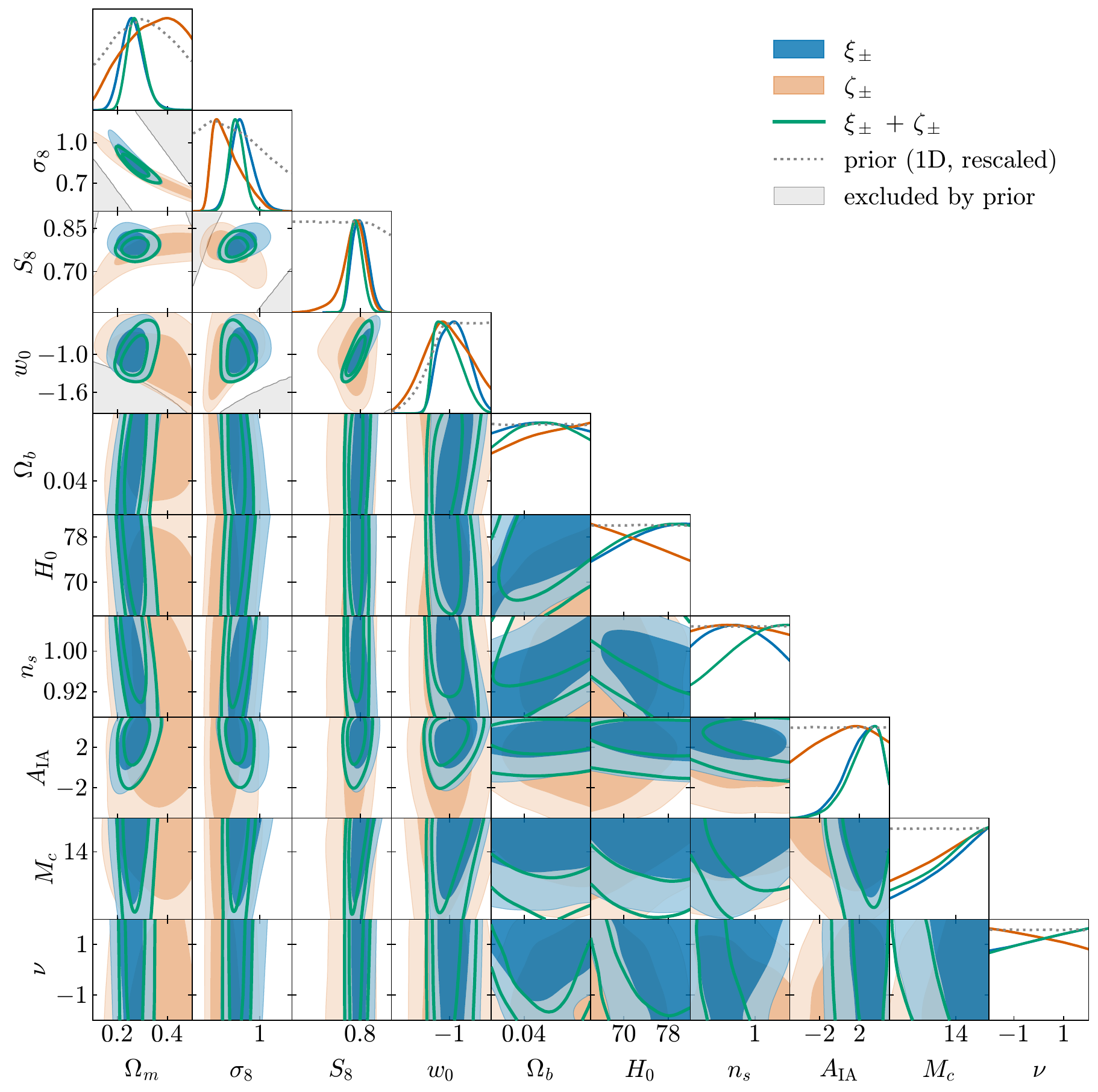}
    \caption{Posterior distributions of all nine sampled parameters and the derived \Seight from the DES Y3 measurement. Blue shows the constraints from \xipm and orange those from \zetapm, while the green unfilled contours show the joint analysis. The grey regions are excluded by the prior, and the dotted lines on the diagonal show the 1D prior marginals.}
    \label{fig:corner_full}
\end{figure}

Figure~\ref{fig:corner_full} shows the complete 1D and 2D marginalised posterior distributions of all nine sampled parameters and the derived \Seight for the \xipm-only (blue), the \zetapm-only (orange) and the joint (green) analysis, complementing the results for the main cosmological parameters of Section~\ref{sec:constraints}. The regions excluded by the prior are shown in grey; they include the restrictions in the \Om--\sigeight and \Om--\wzero planes of Section~\ref{sec:forward_model}. The 1D prior marginals on the diagonal are obtained from a chain sampling the prior and rescaled to the panel height. Table~\ref{tab:map} lists the best-fitting parameters of the three analyses, i.e.~the MAP in the 9-dimensional parameter space (Section~\ref{sec:sbi}), at which the goodness-of-fit tests of Section~\ref{sec:validation} evaluate the forward model.

To quantify which parameters are constrained by the data, we compute the ratio of the 68\% widths of the posterior and of the prior chain. \Ob, \Hzero, \ns, \Mc and $\nu$ are prior-dominated in all three analyses (ratios $\ge 0.90$, with the median of \Mc in the upper half of the prior), and as in Paper~1, the data vector adds little information on these parameters. In the joint analysis, \Aia ($0.49$) and \wzero ($0.57$) are partially constrained, while the well-constrained parameters \Om, \sigeight and \Seight ($0.30$, $0.19$ and $0.11$) are reported in Section~\ref{sec:constraints}. For \xipm alone, the ratios are slightly larger ($0.37$, $0.29$ and $0.14$ for \Om, \sigeight and \Seight, $0.63$ for \wzero and $0.52$ for \Aia), while \zetapm on its own only constrains \sigeight ($0.40$) and \Seight ($0.19$) and leaves \Om, \wzero and \Aia at the prior width.
\begin{table}[H]
    \centering
    \begin{tabular}{|l|ccc|}
        \hline
        \textbf{Parameter} & \xipm & \zetapm & \xipm + \zetapm \\
        \hline
        \Om          & $\phantom{-}0.269$ & $\phantom{-}0.322$ & $\phantom{-}0.243$ \\
        \Ob          & $\phantom{-}0.041$ & $\phantom{-}0.060$ & $\phantom{-}0.054$ \\
        \Hzero       & $79.6$  & $65.8$  & $81.2$  \\
        \ns          & $\phantom{-}0.895$ & $\phantom{-}1.063$ & $\phantom{-}1.068$ \\
        \sigeight    & $\phantom{-}0.869$ & $\phantom{-}0.748$ & $\phantom{-}0.871$ \\
        \wzero       & $-0.72$ & $-1.42$ & $-1.18$ \\
        \Aia         & $\phantom{-}4.02$  & $-1.11$ & $\phantom{-}2.45$  \\
        $\log \Mc$   & $15.00$ & $14.76$ & $14.55$ \\
        $\nu$        & $\phantom{-}0.57$  & $\phantom{-}0.04$  & $\phantom{-}1.99$  \\
        \hline
        \Seight      & $\phantom{-}0.823$ & $\phantom{-}0.775$ & $\phantom{-}0.784$ \\
        \hline
    \end{tabular}
    \caption{Best-fitting parameters of the three main analyses of the DES Y3 measurement: the MAP of the nine sampled parameters, obtained by maximising the log-posterior in the full parameter space (Section~\ref{sec:sbi}), and the derived \Seight.}
    \label{tab:map}
\end{table}

\section{Uncompressed DES Y3 measurements}\label{sec:uncompressed}

\begin{figure}[H]
    \centering
    \includegraphics[width=\linewidth]{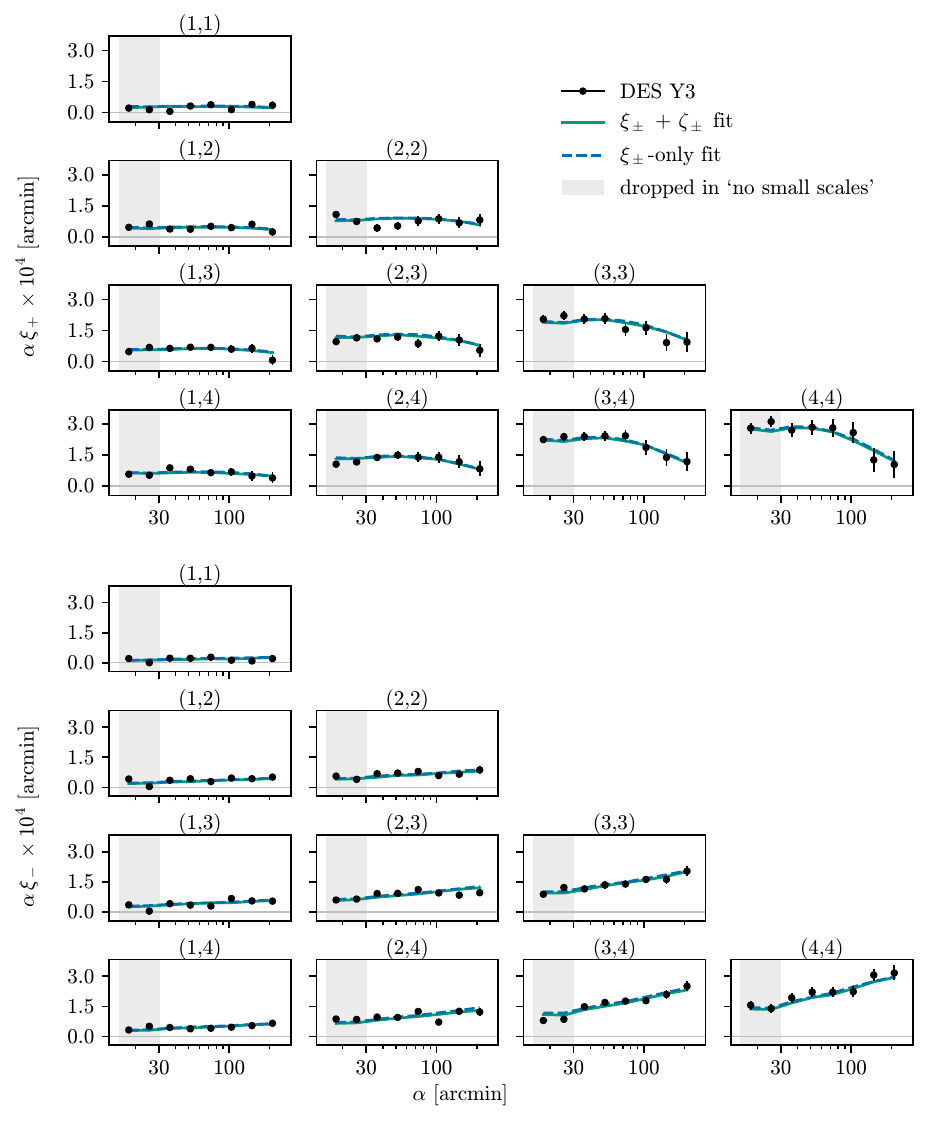}
    \caption{The measured DES Y3 \xip (upper block) and \xim (lower block) for the ten tomographic bin pairs, scaled by the angular separation $\alpha$. The error bars show the scatter of the 16,000 fiducial footprints, and the lines show the posterior predictive means of the joint (green, solid) and the \xipm-only (blue, dashed) analysis. The shaded region marks the scales removed in the `no small scales' variation.}
    \label{fig:datavector_xipm}
\end{figure}

\begin{figure}[H]
    \centering
    \includegraphics[width=\linewidth]{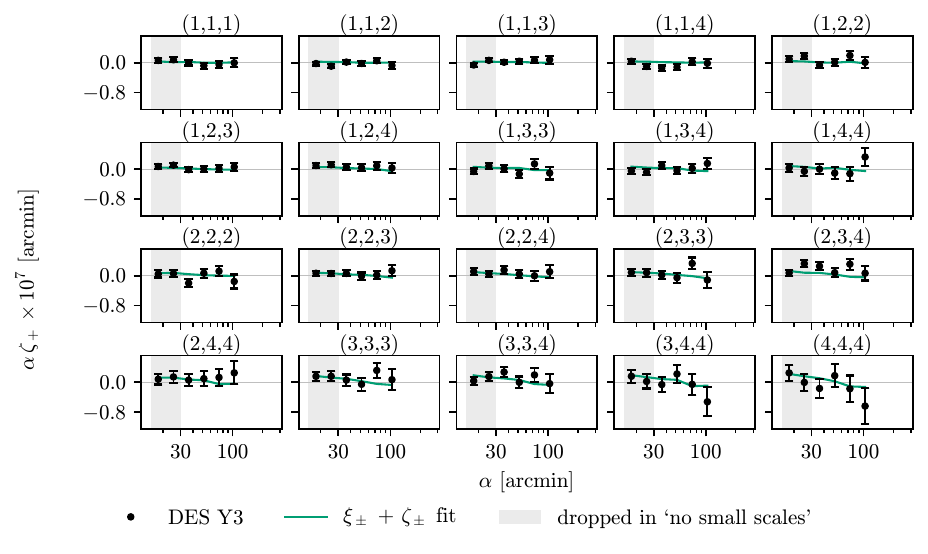}
    \caption{Same as Figure~\ref{fig:datavector_xipm} but for \zetap of the twenty tomographic bin combinations at the aperture radius $\theta_{\rm ap} = 90'$, scaled by the separation $\alpha$ of the local \xipm. The line shows the posterior predictive mean of the joint analysis.}
    \label{fig:datavector_zetap}
\end{figure}

\begin{figure}[H]
    \centering
    \includegraphics[width=\linewidth]{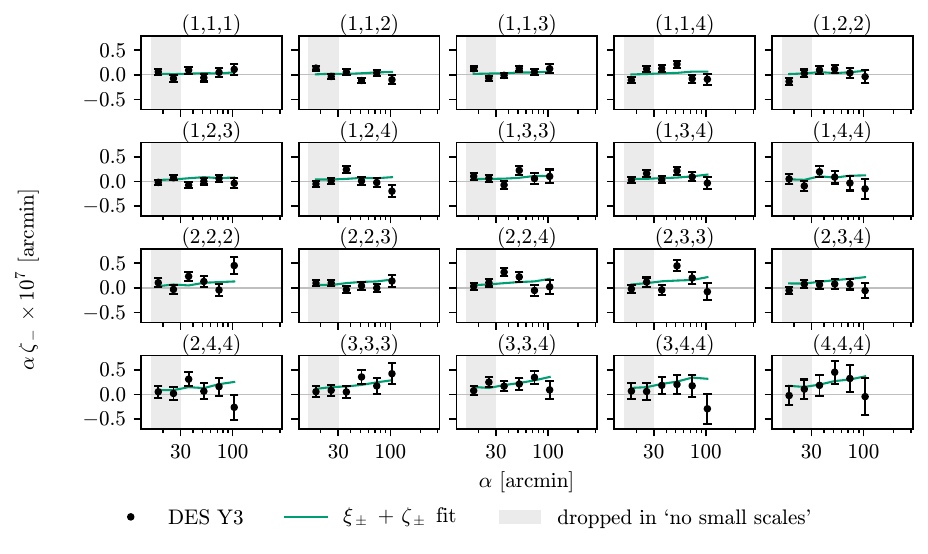}
    \caption{Same as Figure~\ref{fig:datavector_zetap} but for \zetam.}
    \label{fig:datavector_zetam}
\end{figure}

\begin{figure}[H]
    \centering
    \includegraphics[width=\linewidth]{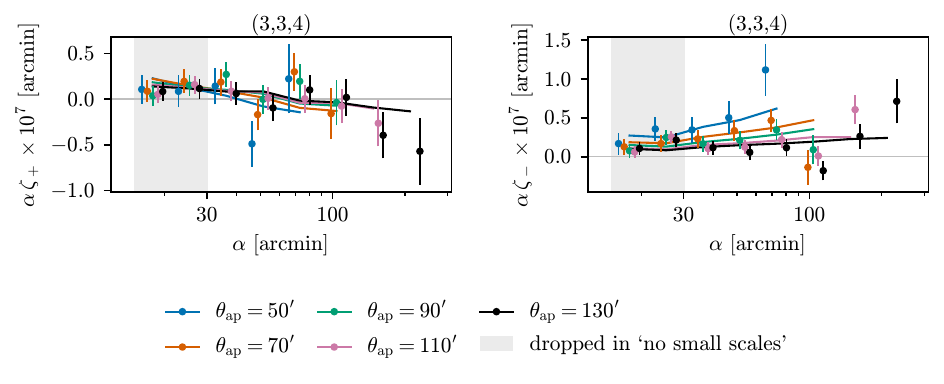}
    \caption{The measured \zetap (left) and \zetam (right) for the five aperture radii, shown for the bin combination $(3,3,4)$, one of the two with the highest signal-to-noise ratio. The points are slightly offset in $\alpha$ for visibility, and the number of angular bins grows with the aperture radius, as $\alpha_{\max} \le 2\theta_{\rm ap} - 5'$. Error bars, lines and shading are as in Figure~\ref{fig:datavector_zetap}.}
    \label{fig:datavector_zeta_apertures}
\end{figure}

Figures~\ref{fig:datavector_xipm}, \ref{fig:datavector_zetap} and \ref{fig:datavector_zetam} show the measured DES Y3 \xipm and \zetapm entering the compression for all tomographic bin combinations, where for the \zetapm combination $(i,j,k)$ \Map is measured in bin $i$ and the local \xipm between bins $j$ and $k$. We show \zetapm for the aperture radius $\theta_{\rm ap} = 90'$, as the other radii look very similar; their dependence on the aperture radius is shown in Figure~\ref{fig:datavector_zeta_apertures} for the combination $(3,3,4)$, which, together with $(2,2,2)$, is one of the two combinations with the highest signal-to-noise ratio of all twenty, with the number of angular bins growing with the radius as $\alpha_{\max} \le 2\theta_{\rm ap} - 5'$. The measurements are scaled by the angular separation, the error bars show the scatter of the 16,000 fiducial footprints (the diagonal of the covariance), and the two smallest angular bins of every correlation, which are removed in the `no small scales' variation, are shaded. In Figure~\ref{fig:datavector_xipm}, the posterior predictive means of the joint and the \xipm-only analysis are nearly indistinguishable, which is the small difference between the two analyses of Table~\ref{tab:constraints} seen in data space.

As we do not have an emulator, we use the posterior predictive mean of the forward-modelled mocks as the model prediction in data space, i.e.~the mean of the 67,500 training and validation data vectors weighted by a Gaussian kernel in parameter space that is centred on the MAP and has the marginal posterior widths of the respective analysis as its widths. The effective number of mocks contributing to this mean, $N_{\rm eff} = (\sum_k w_k)^2 / \sum_k w_k^2$ with the kernel weights $w_k$ \cite{Kisch1965}, is 70 for the joint analysis and 790 and 1627 for the \xipm-only and \zetapm-only posteriors, as their wider kernels include more of the training set. The goodness of fit, $\chi^2 = (x - \mu)^T C^{-1} (x - \mu)$ of the measurement with respect to this prediction, uses the inverse of the fiducial covariance of that analysis corrected by the Hartlap factor \cite{Hartlap2007}, and the percentiles in Table~\ref{tab:constraints} are computed from the same statistic for the MAP cosmology of each test simulation. For the blinded data vector, we found $\chi^2 = 1748$ ($N_{\rm d} = 1440$, 78th percentile) for the joint analysis, $190/160$ (67th) for \xipm and $1548/1280$ (80th) for \zetapm, which is slightly worse than for the unblinded measurement (Table~\ref{tab:constraints}), similar to the MAP percentiles of Appendix~\ref{sec:app_loglike}.

The signal-to-noise ratios of the individual parts of the data vector are $19.2$ for \xip, $29.7$ for \xim, $7.7$ for \zetap and $11.8$ for \zetam. \xim has the higher signal-to-noise ratio of the two, because at a given separation it probes smaller scales than \xip, whose covariance on $15'$--$250'$ is dominated by cosmic variance.

\FloatBarrier
\section{Machine learning parameters}\label{sec:hyperparameters}
\begin{table}[H]
    \centering
    \begin{tabular}{|l|l|}
        \hline
        \textbf{Hyperparameter} & \textbf{Search range} \\
        \hline
        Transforms & $4 - 12$ \\
        Spline bins & $5 - 50$ \\
        Blocks & $\{2, 3, 4\}$ \\
        Features & $\{32, 64, 128, 256, 512\}$ \\
        Learning rate & $[10^{-5}, 10^{-3}]$, log-uniform \\
        Decay factor & $\{0.1, \sqrt{0.1}\}$ \\
        \hline
    \end{tabular}
    \caption{The hyperparameter ranges of the \texttt{optuna} search. Transforms is the number of stacked NSF transformations and Spline bins the number of bins of each spline. Blocks and Features are the number of hidden layers and units per layer of the conditioning networks, and the decay factor reduces the learning rate on plateaus of the validation loss.}
    \label{tab:hyperparameter_ranges}
\end{table}

All density estimators are trained with the \texttt{Adam} optimiser \cite{Kingma2017} and a batch size of 250. The learning rate is multiplied by a tuned decay factor whenever the validation loss has not improved for 5 epochs, and training is stopped once the validation loss has not improved for 20 epochs, after which the model of the best epoch is kept. For every analysis variation, the hyperparameters of the NSF are tuned with \texttt{optuna} \cite{optuna_2019} over the search space listed in Table~\ref{tab:hyperparameter_ranges} (the number of transformations and of spline bins sampled uniformly over the integer ranges, the learning rate log-uniformly, the remaining parameters from the listed discrete choices), using a \texttt{TPESampler} \cite{bergstra2011} with 100 random start-up trials and 300 trials in total, and selecting the model with the highest likelihood on the validation set. The resulting hyperparameters of the three main analyses and the five data-vector variations of Section~\ref{sec:validation} are listed in Table~\ref{tab:hyperparameters}.

\begin{table}[H]
    \centering
    \setlength{\tabcolsep}{4pt}
    \begin{tabular}{|l||r|r|r|r|r|r|}
        \hline
        Analysis & Transforms & Spline bins & Blocks & Features & Learning rate & Decay factor \\
        \hline \hline
        \xipm & 4 & 5 & 2 & 32 & $9.2 \times 10^{-4}$ & $0.1$ \\
        \zetapm & 4 & 7 & 3 & 32 & $9.0 \times 10^{-4}$ & $0.1$ \\
        \xipm + \zetapm & 4 & 5 & 2 & 32 & $9.7 \times 10^{-4}$ & $\sqrt{0.1}$ \\
        \hline
        no small scales & 4 & 6 & 2 & 32 & $9.9 \times 10^{-4}$ & $0.1$ \\
        no BIN1 & 4 & 5 & 2 & 32 & $9.4 \times 10^{-4}$ & $\sqrt{0.1}$ \\
        no BIN2 & 6 & 5 & 2 & 32 & $8.0 \times 10^{-4}$ & $0.1$ \\
        no BIN3 & 4 & 8 & 4 & 64 & $4.1 \times 10^{-4}$ & $0.1$ \\
        no BIN4 & 4 & 5 & 2 & 32 & $8.8 \times 10^{-4}$ & $\sqrt{0.1}$ \\
        \hline
    \end{tabular}
    \caption{The hyperparameters of the best performing NSF for every analysis variation, selected from the 300 \texttt{optuna} trials by the highest likelihood on the validation set. The column names are defined as in Table~\ref{tab:hyperparameter_ranges}.}
    \label{tab:hyperparameters}
\end{table}

\bibliographystyle{JHEP}
\FloatBarrier
\bibliography{biblio.bib}

\end{document}